\documentclass{emulateapj}
\slugcomment{Draft version}

\newcommand\asca{{\it ASCA}}

\newcommand\chandra{{\it Chandra}}

\newcommand\xmm{{\it XMM-Newton}}

\newcommand\s{{\rm~s}}
\newcommand\ks{{\rm~ks}}

\newcommand\hz{{\rm~Hz}}
\newcommand\kev{{\rm~keV}}
\newcommand\ev{{\rm~eV}}
\newcommand\kms{\ifmmode {\rm~km\ s}^{-1} \else ~km s$^{-1}$\fi}
\newcommand\Hunit{\ifmmode {\rm~km\ s}^{-1}\ {\rm Mpc}^{-1}
        \else ~km s$^{-1}$ Mpc$^{-1}$\fi}
\newcommand\ctssec{\ifmmode {\rm~count\ s}^{-1} \else ~count s$^{-1}$\fi}
\newcommand\ergsec{\ifmmode {\rm~erg\ s}^{-1} \else
        ~erg s$^{-1}$\fi}
\newcommand\funit{\ifmmode {\rm~erg\ s}^{-1}\;{\rm cm}^{-2} \else
        ~ergs s$^{-1}$ cm$^{-2}$\fi}
\newcommand\phflux{\ifmmode {\rm~photon\ s}^{-1}\;{\rm cm}^{-2}
        \else   ~photon s$^{-1}$ cm$^{-2}$\fi}
\newcommand\efluxA{\ifmmode {\rm~erg\ s}^{-1}\;{\rm cm}^{-2}\;{\rm
        \AA}^{-1} \else ~erg s$^{-1}$ cm$^{-2}$ \AA$^{-1}$\fi}
\newcommand\efluxHz{\ifmmode {\rm~erg\ s}^{-1}\;{\rm cm}^{-2}\;{\rm
        Hz}^{-1} \else ~erg s$^{-1}$ cm$^{-2}$ Hz$^{-1}$\fi}
\newcommand\cc{\ifmmode {\rm~cm}^{-3} \else cm$^{-3}$\fi}
\newcommand\FWHM{\ifmmode {\rm~FWHM} \else ${\rm~FWHM}$\fi}
\newcommand\Msun{\ifmmode M_{\odot} \else $M_{\odot}$\fi}
\newcommand\Lsun{\ifmmode L_{\odot} \else $L_{\odot}$\fi}
\newcommand\ltsim{\raisebox{-.5ex}{$\;\stackrel{<}{\sim}\;$}}

\newcommand\hbeta{\ifmmode {\rm H}\beta \else H$\beta$\fi}
\newcommand\Kalpha{\ifmmode {\rm K}\alpha \else K$\alpha$\fi}
\newcommand\nh{\ifmmode N_{\rm H} \else N$_{\rm H}$\fi}
\usepackage{graphicx}
\usepackage{natbib}
\begin{document}

\title{X-ray emission from  active galactic nuclei with intermediate mass black holes}

\author{G. C. Dewangan\altaffilmark{1}, S. Mathur\altaffilmark{2},
  R. E. Griffiths\altaffilmark{1}, \& A. R. Rao\altaffilmark{3}} 
\altaffiltext{1}{Department of Physics, Carnegie Mellon University,
  5000 Forbes Avenue, Pittsburgh, PA 15213 USA; {\tt email:
    gulabd@cmu.edu, griffith@astro.phys.cmu.edu}
\altaffilmark{2}{Astronomy Department, The Ohio State University, 140
  West 18th Avenue, Columbus, OH 43210, USA}
\altaffilmark{3}{Department of Astronomy and Astrophysics, Tata
  Institute of Fundamental Research, Mumbai 400005, India} }

\begin{abstract}
  We present a systematic X-ray study of eight active galactic nuclei
  (AGNs) with intermediate mass black holes ($M_{BH} \sim 8-95\times
  10^4{\rm~M_{\odot}}$) based on 12 \xmm{} observations. The sample
  includes the two prototype AGNs in this class -- NGC~4395 and POX~52
  and six other AGNs discovered with the Sloan Digitized Sky Survey. These
  AGNs show some of the strongest X-ray variability with the
  normalized excess variances being the largest and the power density
  break time scales being the shortest observed among radio-quiet
  AGNs. The excess variance -- luminosity correlation appears to
  depend on both the BH mass and the Eddington luminosity
  ratio. 
%POX~52, which has the highest Eddington luminosity in our
%  sample, also has the highest normalized excess variance.  
The break
  time scale -- black hole mass relations for AGN with IMBHs are
  consistent with that observed for massive AGNs. 
%The BH mass of NGC~4395 is likely to be lower than that estimated from the
%  reverberation mapping.  
  We find that the FWHM of the H$\beta$/H$\alpha$ line is uncorrelated
  with the BH mass, but shows strong anticorrelation with the
  Eddington luminosity ratio.  Four AGNs show clear evidence for soft
  X-ray excess emission ($kT_{in} \sim 150-200\ev$). X-ray spectra of
  three other AGNs are consistent with the presence of the soft excess
  emission. NGC~4395 with lowest $L/L_{Edd}$ lacks the soft excess
  emission.  Evidently small black mass is not the primary driver of
  strong soft X-ray excess emission from AGNs.  The X-ray spectral
  properties and optical-to-X-ray spectral energy distributions of
  these AGNs are similar to those of Seyfert 1 galaxies. The observed
  X-ray/UV properties of AGNs with IMBHs are consistent with these
  AGNs being low mass extension of more massive AGNs; those with high
  Eddington luminosity ratio looking more like narrow-line Seyfert 1s
  while those with low $L/L_{Edd}$ looking more like broad-line
  Seyfert 1s.
\end{abstract}


\section{Introduction}
The existence of astrophysical black holes in two mass ranges, stellar
mass black holes with $\sim 10M_{\odot}$ in X-ray binaries and
supermassive black holes (SMBHs) with masses in the range of $\sim10^6
- 10^{9}M_{\odot}$ at the centers of active galaxies, is well
supported by observations. Intermediate mass black holes (IMBHs),
bridging the gap between the stellar mass and supermassive black
holes, remain relatively unexplored.  There was no observational
evidence for such black holes for many years.  Recent optical
and X-ray observations have revived the possibility of existence of
IMBHs.
%The discovery of a large number of
%off-nuclear, ultra-luminous X-ray sources (ULXs) in nearby galaxies
%with luminosities in the $\sim 10^{39} - 10^{41}{\rm~ergs~s^{-1}}$
%range suggests existence of IMBH in the $\sim 100-1000M_{\odot}$
%\citep[e.g.,][]{2004PThPS.155...27M,2004IJMPD..13....1M}.  
There are indirect evidence for IMBHs with masses of $\sim
10^4-10^6M_{\odot}$ at the center of some galaxies based on radiative
signatures or dynamical measurements
\citep{2003ApJ...588L..13F,2004ApJ...607...90B,
  2004ApJ...610..722G,2007arXiv0707.2617G}. The AGNs with IMBHs form
an important class for a number of reasons.  First, the growth of
SMBHs in Seyfert galaxies and quasars is still not understood.  The
nature of seed black hole (BH) is the major challenge to any
cosmological BH growth model.  Observations of IMBHs at the dynamical
centers of nearby galaxies would provide significant constraints on
the nature of seed black holes for cosmological growth models.  The
$M_{BH}-\sigma_{*}$ relation strongly suggests co-evolution of
galaxies and BHs, but it is not known if this relation is established
early in the evolution process. If this relation is established late
in the galaxy evolution, as suggested by \cite{2003ApJ...593...56D}, then
we should see deviations in the $M_{BH}-\sigma_{*}$ relation in a
population of IMBHs that is not yet fully grown. In this growing
phase, the accretion process and the disk-corona geometry around the
central IMBH could be markedly different than that around SMBHs in
luminous AGNs. Therefore, it is crucial to study accretion onto the
least massive central BHs in galaxies. AGNs with IMBHs also provide
the missing link between the stellar mass and SMBHs and enable us to
test if the accretion physics is the same at all scales.
%They fill the gap between the stellar mass
%and supermassive black holes and may provide a direct link between the
%accretion processes of stellar mass black hole bineries (BHBs) and AGNs. They provide the direct link
%between the temporal characteristics of BHBs and AGNs.  
They will also
help to understand the accretion physics as it would be easier to
isolate certain characteristics arising solely from an extreme
physical parameter, in this case, low mass BH. AGNs with IMBHs may
reveal new accretion phenomena depending on the mode of accretion at
low BH masses.  It is also possible that they pass through an active
phase of accretion and contribute to the observed background radiation
at some level, particularly to the X-ray background. The mergers of
intermediate mass black holes with masses $10^5M_{\odot}$ are expected
to provide strong signals of gravitational waves for the {\it Laser
  Interferometric Space Antenna} \citep[e.g.,][]{2002MNRAS.331..805H}.

The two prototypes examples of
IMBH AGNs are the late-type spiral galaxy NGC~4395
\citep{2003ApJ...588L..13F} and the dwarf elliptical galaxy POX~52
\citep{2004ApJ...607...90B}. \cite{2004ApJ...610..722G} discovered 19
such AGNs using the Sloan Digital Sky Survey (SDSS). They have
increased their sample by an order of magnitude to 174 using the
fourth data release of the SDSS \citep{2007arXiv0707.2617G}.
%The well known AGN with IMBH NGC~4395 is the prototype of this
%class. It is the least luminous known Seyfert 1 galaxy that was
%discovered by \cite{1989ApJ...342L..11F}.  
NGC~4395 has the emission properties of a type 1 AGN, with broad
optical and UV emission lines and a point-like hard X-ray source. Its
X-ray emission is extremely rapidly variable \citep[see
e.g.,][]{2005MNRAS.356..524V}. \cite{2003ApJ...588L..13F} argued that
the true mass of the black hole is likely to be $\sim
10^4-10^5M_{\odot}$, while reverberation mapping of NGC~4395 has
provided a BH mass of $(3.6\pm1.1)\times10^{5}M_{\odot}$
\citep{2005ApJ...632..799P}.  Another recently discovered AGN with an
IMBH is POX~52, a dwarf elliptical galaxy at $z=0.021$. It was
discovered in an objective-prism search for emission line objects by
\cite{1981A&AS...44..229K} who also noted its star-like appearance in
the Palomar Sky Survey images.  Follow-up imaging and spectroscopic
observations by \cite{1987AJ.....93...29K} revealed a dwarf galaxy
with an AGN spectrum. The object was classified as a Seyfert 2 galaxy
by \citet{1981A&AS...44..229K} based on the flux-ratios of narrow
emission lines although a weak broad component of the H$\beta$ line
was seen with a full width at half maximum, FWHM $\sim 840\kms$.  The
galaxy is indeed quite small with a diameter $\sim 6.4{\rm~kpc}$ and
its absolute magnitude is only $M_{V} = -16.9$. The luminosity of the
nucleus is comparable to that of a mild Seyfert 2 nucleus but the
ratio of nuclear to host-galaxy luminosity is 1.1, a typical value for
quasars \citep{1987AJ.....93...29K}.  \cite{2004ApJ...607...90B}
obtained a high quality optical spectra of POX~52 at the Keck that
revealed an emission line spectrum very similar to that of the dwarf
Seyfert 1 galaxy NGC~4395, with broad components to the permitted
lines and reclassified POX~52 as a Seyfert 1 galaxy.  This galaxy has
a central velocity dispersion of $36\kms$, which yields a mass of
$1.4\times 10^{5}{\rm~M_{\odot}}$, again consistent with that derived
from the H$\beta$ line width \citep{2004ApJ...607...90B}.
%\cite{2004ApJ...610..722G} discovered 19
%such AGNs using the Sloan Digital Sky Survey (SDSS).  Recently,
%\cite{2007arXiv0707.2617G} have increased their sample to 174 using
%the 4th SDSS data release.

\begin{table*}
\centering
  \caption{General properties of AGN with intermediate mass black
    holes observed with \xmm{} \label{t1}}
\begin{tabular}{lcccccccc} \tableline\tableline
  Object  &  \multicolumn{2}{c}{Position} & Redshift & Galactic $N_H$ & FWHM(${\rm km~s^{-1}}$)
  & $M_{BH}$(${\rm M_{\odot}}$) & $\frac{L_{bol}}{L_{Edd}}$ & Reference \\
           & $\alpha$(J2000.0) & $\delta$(J2000.0) & ($z$) & ($10^{20}{\rm~cm^{-2}}$) & ($H\alpha$ or $H\beta$) &    &  & \\ \tableline
NGC~4395 &  12 25 48.93 & +33 32 47.8 & $0.001$ & $1.85$ & $1500$ & $3.6\times10^{5}$ & $1.2\times10^{-3}$ & (1) \\
POX~52   &  12 02 56.90 & --20 56 03.0  & $0.021$  & $3.85$ & $760$ & $1.6\times10^{5}$ & $0.5-1$ & (2)  \\
SDSS~J010712.03+140844.9 & 01 07 12.03 & +14 08 44.9 & $0.0768$ & $3.44$ & $830$ & $7.2\times10^{5}$ & $1.24$  & (3)  \\
SDSS J024912.86-081525.6 & 02 49 12.86 & --08 15 25.6 & $0.0295$ & $3.67$ & $732$ &  $8.3\times10^{4}$ & $0.71$ & (3) \\
SDSS J082912.67+500652.3 & 08 29 12.67 & +50 06 52.3 & $0.0434$ & $4.08$ & $870$ & $5.2\times10^{5}$ & $0.95$ & (3) \\
SDSS J114008.71+030711.4 & 11 40 08.71 & +03 07 11.4 & $0.0811$ & $1.91$ & $591$ & $5.9\times10^{5}$ & $2.98$ & (3) \\
SDSS J135724.52+652505.8 & 13 57 24.52 &  +65 25 05.8 & $0.106$ & $1.36$ & $872$ &  $9.5\times10^{5}$ & $1.21$ & (3)  \\
SDSS J143450.62+033842.5 & 14 34 50.71 & +03 38 40.4 & $0.0284$ & $2.43$ & $1089$ & $2.0\times10^{5}$ & $0.33$ & (3) \\ \tableline
\end{tabular}
References: (1) \citet{2005ApJ...632..799P}, (2) \cite{2004ApJ...607...90B} (3) \cite{2004ApJ...610..722G}.
\end{table*}

 AGNs with IMBHs are rare compared to luminous AGNs.  One explanation
is that they are short-lived. Presumably, all AGNs grow fast in their
early phase until their BH mass increases to
$10^6{\rm~M_{\odot}}$.  During this fast-growth phase, the accretion rate
may be close to Eddington or super-Eddington. The relative accretion
rates of $\sim 1$ for the SDSS AGNs with IMBH, derived by \cite{2004ApJ...610..722G} from the optical data, favors the above scenario. However, all
AGNs with IMBH are not accreting at high rates e.g., NGC~4395 is
underlumnious for its BH mass with $L/L_{Edd} \sim 1.2\times 10^{-3}$
\citep{2005ApJ...632..799P} or at most $0.2$ \citep{2005MNRAS.356..524V}
In fact, very little is known about the accretion process onto IMBHs
in AGNs. The only AGN that has been well studied in X-rays is
NGC~4395. \cite{2000MNRAS.318..879I} presented the ASCA
spectrum that showed a power-law continuum of photon index
$\Gamma=1.7\pm0.3$ with a Fe K line marginally detected at $\sim
6.4\kev$. The soft-X-ray emission below $3\kev$ is strongly attenuated
by absorption. The X-ray spectrum in this absorption band showed a
dramatic change in response to the variation in continuum luminosity.
 \cite{2000MNRAS.318..879I} concluded that the nuclear source of NGC~4395 is
consistent with a scaled-down version of higher-luminosity Seyfert
nuclei.  \cite{2005MNRAS.356..524V} have studied the
exceptional X-ray variability of NGC~4395. The variations observed are
among the most violent seen in an AGN to date, with the fractional rms
amplitude exceeding $100\%$ in the softest band. 
%The amplitude of the
%variations seems intrinsically higher in NGC 4395 than most other
%Seyfert galaxies, even after accounting for the differences in BH
%masses. The origin of this difference is not clear, but it is unlikely
%to be a high accretion rate \citep{2005MNRAS.356..524V}.

\cite{2007ApJ...656...84G} have performed snapshot ($5\ks$ exposure)
\chandra{} ACIS X-ray observations of IMBH AGNs discovered with
SDSS. They detected 8 of the 10 AGNs with IMBH with a significance
$\ge 3\sigma$. They measured the $0.5-2\kev$ photon indices in the
range of $1-2.7$, consistent with that for luminous AGNs, implying
that the BH mass is not the fundamental driver for the soft X-ray
spectral shape. However, this result is not robust as some of the
measurements are based on hardness ratios. Moreover, with $5\ks$
\chandra{} exposures, it was not possible to make detailed spectral
and temporal analysis of any of the AGNs.

\begin{table*}
\centering
\caption{Log of \xmm{} observations \label{t2}}
\begin{tabular}{lcccc} \tableline\tableline
Object &  ObsID & Start date & pn rate & pn exposure  \\ 
       &        &            & (${\rm~counts~s^{-1}}$) & (${\rm~ks}$) \\ \tableline
NGC~4395 & 0112522001(1) & 2002-06-12 & 	 \\
         & 0112521901(2) & 2002-05-31 &  $0.56\pm0.007$ & $12.5$ \\
         & 0142830101(3) & 2003-11-30 & $1.13\pm0.004$ & $93.0$  \\
         & 0112522701(4) & 2003-01-03 & $1.14\pm0.014$ & $5.9$\\
POX~52   & 0302420101(1) & 2005-07-08 & $0.05\pm0.001$ & $81.1$ \\
J0107+1408 & 0305920101(1) & 2005-07-22  & $0.27\pm0.003$ & $24.9$ \\
J0249--0815 & 0303550101(1) & 2006-02-16 & $0.23\pm0.008$ & $4.0$ \\
J0829+5006 & 0303550301(1) & 2006-03-27 &   \\
        & 0303550801(2) & 2006-03-28 &  \\
        & 0303550901(3) & 2006-04-26 & $0.95\pm0.02$ & $2.3$ \\
        & 0303551001(4) & 2006-04-26 &  \\        
J1140+0307 & 0305920201(1) & 2005-12-03 & $0.66\pm0.004$ & $35.1$ \\
J1357+6525 & 0305920301(1) & 2005-04-04 & $0.45\pm0.005$ & $19.2$  \\
       & 0305920501(2) & 2005-05-16 & $0.55\pm0.01$ & $1.9$ \\
       & 0305920601(3) & 2005-06-23 & $0.42\pm0.006$ & $12.0$  \\ 
J1434+0338 & 0305920401(1) & 2005-08-18 & $0.10\pm0.002$ & $22.0$ \\ \tableline
\end{tabular}
\end{table*}

In this paper, we present X-ray temporal and spectral study of a
sample of 8 AGNs with IMBHs.  These AGNs with $M_{BH} < 10^6M_{\odot}$
were selected based on the availability of \xmm{} data. In
Table~\ref{t1}, we list these AGNs along with their general
properties.  The least massive AGN in our sample has only $M_{BH} \sim
83000M\odot$. The accretion rates of these AGNs, estimated from
optical observations, has a large range ($L_{bol}/L_{Edd} \sim 10^{-3}
- 3$), suggesting a large diversity in the X-ray properties of these
AGNs. We also present optical/UV emission observed with \xmm{} optical
monitor (OM).  We describe \xmm{} observations and data reduction in
Section 2. In Section 3, we present temporal analysis of the EPIC-pn
data, followed by spectral analysis in Section 4. We present the
analysis of OM data and derive optical-to-X-ray spectral indices
($\alpha_{ox}$) in Section 5.  We discuss the results in Section 6,
followed by a summary of our study in Section 7.

We assume the following cosmological parameters to calculate
distances; $H_0 = 71{\rm~km~s^{-1}~Mpc^{-1}}$, $\Omega_m = 0.27$, and
$\Omega_{\Lambda} = 0.73$. In the following, we abbreviate the names
of the AGNs with IMBHs discovered with the SDSS e.g.,
we mention SDSS~J082912.67+500652.3 as J0829+5006.

\section{Observation \& Data Reduction}
All the eight AGNs with IMBHs presented here were observed with \xmm{}
between 2002 May and 2006 April. The European Photon Imaging Cameras
(EPIC) pn \citep{2001A&A...365L..18S} and MOS \citep[MOS1 and MOS2;][]{2001A&A...365L..27T} were operated in the imaging mode during all the observations.
Table~\ref{t2} lists the details of the X-ray observations. NGC~4395,
J0829+5006 and J1357+6525 were observed
multiple times. The observation data files were processed to produce
calibrated event lists using the Science Analysis System (SAS v7.0.0).
Examination of the background rate above $10\kev$ showed that the
first observation of NGC~4395 on 2002 June and the first, second and
fourth observations of J0829+5006 were completely
swamped out by the flaring particle background and therefore these
observations were discarded from further study. The rest of
observations were either clean or partly affected by the particle
background and the intervals of high particle background were
discarded from spectral analysis. We extracted the source spectra
using the good EPIC-pn events in circular regions of radii in the $25-40\arcsec$
centered at the source position. We used event with pattern $0-4$
(single and double pixel events) for all observations except for the long
observation of NGC~4395. We
used only the single pixel events for the long $\sim 100\ks$
observation of NGC~4395.
The background spectra were similarly
extracted from nearby circular regions free of sources. Spectral
response files were generated using the SAS tasks `rmfgen' and
`arfgen'. In Table~\ref{t1}, we have listed the `cleaned' exposure
obtained from the EPIC-pn spectra. For temporal analysis we considered
the full lengths of the observations after correcting for the varying
background rate. The source and background lightcurves were extracted
using EPIC-pn events with pattern $0-12$ and similar extraction
regions described above. The source lightcurves were corrected for
background contributions and telemetry dropouts.

The optical/UV monitor, co-aligned with the X-ray telescopes, was
operated in the imaging mode utilizing one or more optical/UV filters,
thus providing simultaneous optical/UV/X-ray coverage. The OM data
were processed with the SAS task `omichain'.

\begin{figure*}
\centering \includegraphics[width=17cm]{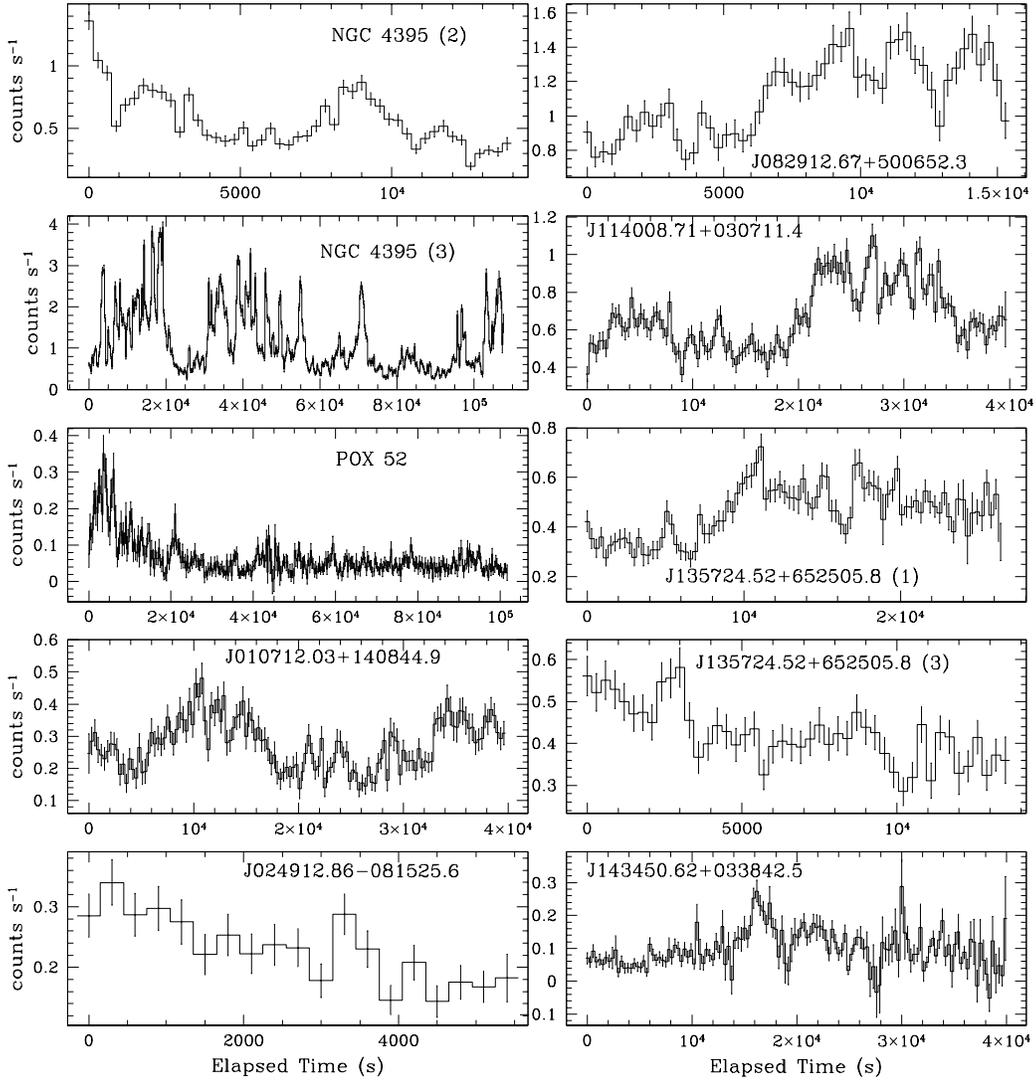}
\caption{\xmm{} X-ray lightcurves of AGNs with IMBH derived from the
  EPIC-pn data with $300\s$ bins.}  
\label{f1}
\end{figure*}

\section{Temporal Analysis}
%The X-ray lightcurves of all AGNs with IMBHs were extracted from the
%EPIC-pn data in the full energy band of $0.2-12\kev$ with $300\s$ bins
%using circular regions with radii in the range of
%$25-45\arcsec$. Background lightcurves were also extracted using
%appropriate circular regions close to but free of source events. The
%source lightcurves were corrected for background contribution and
%telemetry dropouts.  
The background-corrected EPIC-pn lightcurves of the eight AGNs are
shown in Figure~\ref{f1} with $300\s$ bins.  There are two lightcurves for each of
NGC~4395 and J1357+6525 extracted from different
observations. All AGNs exhibit strong and rapid X-ray variability. Individual AGNs show
different amplitudes of variability. The most remarkable variability
is seen in the lightcurve of NGC~4395. X-ray emission from Pox~52 is also highly
variable.  The $0.2-10\kev$ band light curve shows a trough-to-peak
variations by a factor of $\sim10$. There are several rapid
variability events e.g., roughly one event in every $10000\s$,
throughout the observations. The count rate is found to increase by a
factor $\ga 2$ in $\sim2500\s$ in these events. The most dramatic
event is seen near the start of the observation. The count rate
increased by a factor of $\sim2$ and decreased immediately by a factor
$\sim 3$ in an interval of about $\sim 5400\s$. To quantify the variability
properties of individual AGNs, we have measured the intrinsic source
variability expressed in terms of the excess variance
($\sigma^2_{XS}$) i.e., the variance after subtracting the
contribution expected from measurement errors, the normalized excess
variance ($\sigma^2_{NXS} = \sigma^2_{XS}/\bar{x}^2$ with $\bar{x}$ is
the mean count rate) and the fractional root mean square (rms)
variability amplitude ($F_{var} = \sqrt{\sigma^2_{NXS}}$) (see Vaughan
et al. 2003 and references therein). These quantities are listed in
Table~\ref{t3} for each lightcurve. NGC~4395 in the third observation
and POX~52 show the largest fractional variability amplitudes of
$65.6\pm4.0\%$ and $73.6\pm9.3\%$, respectively.  The fractional
variability amplitude of NGC~4395 calculate here is similar within
errors to that calculated by \cite{2005MNRAS.356..524V} who have performed
a detailed analysis of temporal characteristics of NGC~4395 using the
third \xmm{} observation.  \cite{2005MNRAS.356..524V} also measured the the
soft ($0.2-0.7$) band $F_{var}$ to be in excess of unity, making
NGC~4395 the most variable AGN. In the case of POX~52, some
contribution to the large $F_{var}$ may arise from incorrect
background correction as the long observation was partly affected with
flaring background (see below). 

\begin{table*}
\centering
  \caption{General properties of the EPIC-pn lightcurves of AGNs with IMBH \label{t3}}
\begin{tabular}{lcccccccc}   \tableline\tableline
Object & Data & Bin size & Number  & mean rate & rms & Excess variance & Normalized Excess
 & $F_{var}$ \\
       &      & ($\s$) & of bins  & (${\rm count~s^{-1}}$) &  & ($\sigma_{XS}^2$) &
       variance ($\sigma_{NXS}^2$) & ($\%$) \\ \tableline
NGC~4395 & (2) & $300$ & $47$ & $0.57\pm0.047$ & $0.22\pm0.045$ &
$0.05\pm0.02$ & $0.15\pm0.06$ & $39.0\pm7.8$  \\
         & (3) & $300$ &  $360$ & $1.17\pm0.071$ & $0.77\pm0.057$ &
       $0.59\pm0.088$ &    $0.43\pm0.053$ & $65.6\pm4.0$  \\
POX~52   & (1) & $300$ & $340$ & $0.06\pm0.02$ & $0.05\pm0.012$ &
$0.002\pm0.001$ & $0.54\pm0.14$ & $73.6\pm9.3$  \\
J010712.03+140844.9 & (1) & $300$ & $133$ & $0.27\pm0.036$ &
$0.07\pm0.02$ & $0.005\pm0.003$ & $0.06\pm0.04$ & $24.7\pm7.5$ \\
J024912.86--081525.6 & (1) & $300$ & $19$ & $0.23\pm0.03$ &
$0.04\pm0.03$ & $0.002\pm0.002$ & $0.04\pm0.05$ & $19.7\pm12.7$ \\
J082912.67+500652.3 & (3) & $300$ & $52$ & $1.11\pm0.08$ &
$0.21\pm0.08$ & $0.04\pm0.033$ & $0.03\pm0.02$ & $18.6\pm7.2$ \\
J114008.71+030711.4 & (1) & $300$ & $133$ & $0.67\pm0.05$ &
$0.16\pm0.04$ & $0.03\pm0.013$ & $0.06\pm0.03$ & $24.4\pm5.8$ \\
J135724.52+652505.8 & (1) & $300$  & $89$ & $0.46\pm0.05$ &
$0.09\pm0.03$ & $0.009\pm0.007$ & $0.04\pm0.03$ & $20.5\pm7.4$ \\
                    &  (3) & $300$ & $46$ & $0.42\pm0.04$ &
                    $0.06\pm0.03$ & $0.003\pm0.004$ & $0.02\pm0.02$ &
                    $13.9\pm8.4$ \\
J143450.62+033842.5 & (1) & $300$ & $134$ & $0.10\pm0.04$ &
$0.04\pm0.02$ & $0.002\pm0.001$ & $0.20\pm0.11$ & $44.6\pm13.0$ \\ \tableline
\end{tabular}
\end{table*}

\begin{figure}
\centering
\includegraphics[width=10cm]{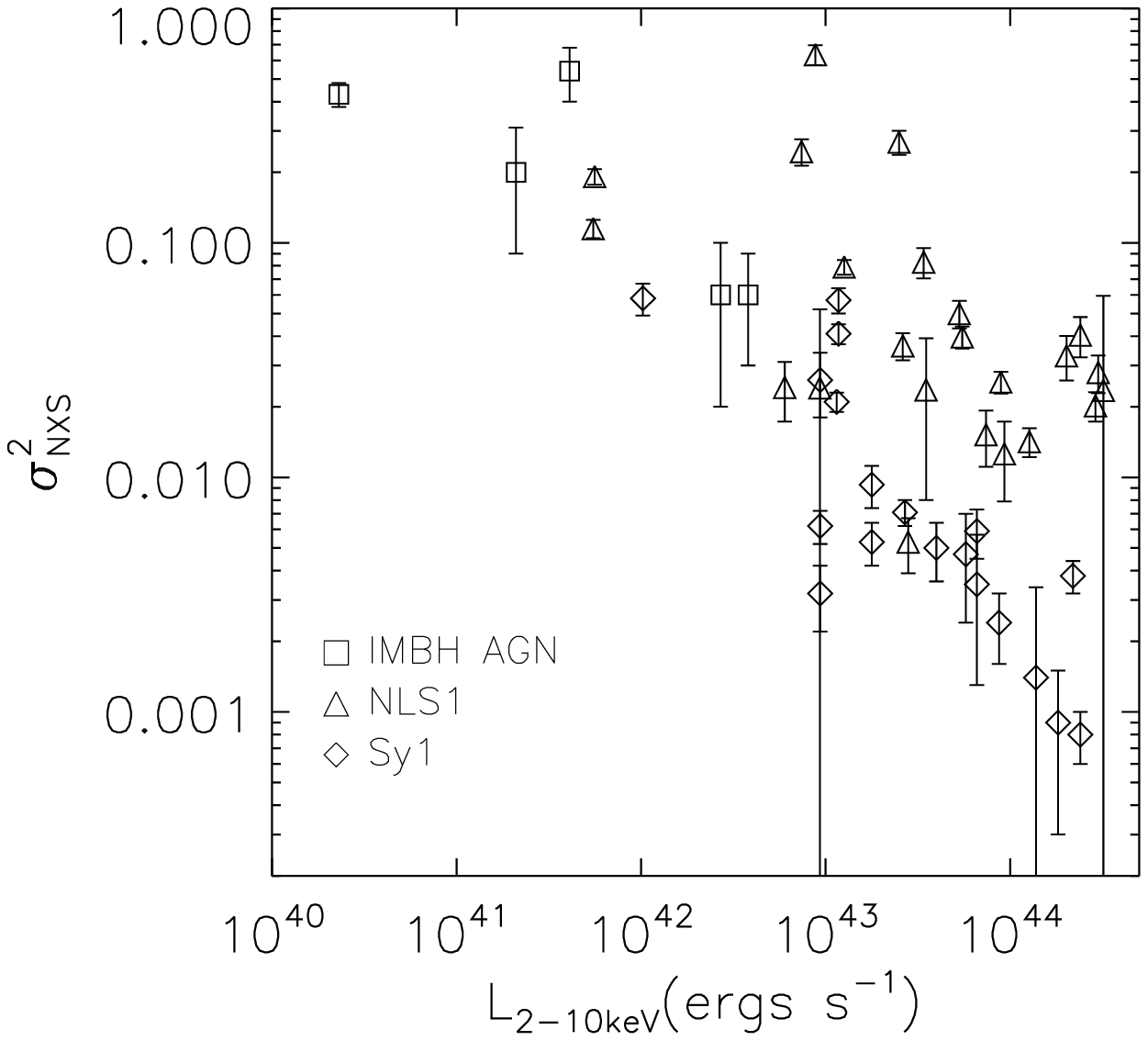}
\caption{Normalized excess variance and $2-10\kev$ X-ray luminosity
  for AGNs with IMBH, NLS1 galaxies \citep{1999ApJS..125..297L} and Syfert 1
  galaxies \citep{1997ApJ...476...70N}}.
\label{exc_var_lum}
\end{figure}

One of the objectives of this paper is to compare the variability
characteristics of AGNs with IMBHs with that of NLS1s and broad-line
Seyfert 1 galaxies (BLS1s). 
%Instead of making time series analysis of
%samples of NLS1s and BLS1s, we have used published
%results.
 \cite{1999ApJS..125..297L} presented a comprehensive X-ray
variability of a sample of NLS1s based on \asca{}
observations. \cite{1997ApJ...476...70N} presented a systematic time
series analysis of Seyfert 1 galaxies consisting mostly of BLS1s using
\asca{} data. It is preferable to use results based on \xmm{}
observations. However, we did not find systematic \xmm{} study of samples of NLS1s
and BLS1s, similar to that performed by \cite{1999ApJS..125..297L} and
\cite{1997ApJ...476...70N}. Therefore, we used results from
\cite{1999ApJS..125..297L} for NLS1 and \cite{1997ApJ...476...70N} for
BLS1s with the exclusion of two NLS1s NGC~4051 and Mrk~335.  In
Figure~\ref{exc_var_lum}, we have plotted the normalized excess
variance as a function of $2-10\kev$ luminosity for AGNs with IMBHs,
NLS1s and BLS1s.  As pointed out by \cite{1993MNRAS.261..612P} and
\cite{1999ApJS..125..297L}, the normalized excess variance depends on
the length of observation. The normalized excess variance is the
integration of the power density spectrum over a certain frequency
range implied by the bin size and the length of the observation. The
rednoise nature of the AGN variability means a strong dependance of
excess variance on the length of the observation.  Therefore, care
must be taken in interpreting Fig.~\ref{exc_var_lum}. Some of AGNs with IMBH
were observed for a short duration. To minimize the effect of
observation length, we have plotted only those five AGNs with IMBHs
that were observed for at least $30\ks$, thus making the length of
observations similar to that for the NLS1s in
\cite{1999ApJS..125..297L} and BLS1s in \cite{1997ApJ...476...70N}.
The prototype of AGNs with IMBH NGC~4395 lies at the low luminosity
end in Fig.~\ref{exc_var_lum}. The excess variance of NGC~4395 is comparable
to some of the most variable NLS1s. \citet{2005MNRAS.356..524V} have
noted that the variations in the soft X-ray emission from NGC~4395
are among the most violent seen in an AGN to date, with the fractional
variability amplitude exceeding $100\%$. Previously,
\cite{1997ApJ...476...70N} and \cite{1999ApJS..125..297L} have noted
that the excess variance is inversely correlated with the X-ray
luminosity though with a flatter slope of $\sim -0.3$ for NLS1s compared
with the slope of $\sim -1$ for BLS1s. 
The five AGNs with IMBHs follow the general trend of anticorrelation between $sigma^2$ and $L_{2-10}\kev$ of Seyfert galaxies. While three of them follow the anticorrelation for BLS1, one (POX 52) is clearly in the NLS1 zone while NGC 4395 lies below the extrapolation of the BLS1 trend.
%The five AGNs with IMBHs appear
%to fall nicely in the correlation between excess variance and luminosity
%known for BLS1s and NLS1s.

%Figure~\ref{f2} compares the net EPIC-pn lightcurve of NGC~4395 and
%the background variations.

\subsection{Power density spectra}
Only two AGNs, NGC~4395 and POX~52, were observed with long $\sim
100\ks$ exposures that allow us to probe power density spectra (PDS)
over a broad range of frequency. \cite{2005MNRAS.356..524V} have
already derived the PDS of NGC~4395 using the long observation that
showed a clear break from a flat spectrum ($\alpha \sim 1$) to a
steeper spectrum ($\alpha \sim 2$) at a frequency $0.5-3.0 \times
10^{-3}{\rm~Hz}$.  Here we present the PDS for POX~52.

\begin{figure}
  \centering \includegraphics[width=9cm]{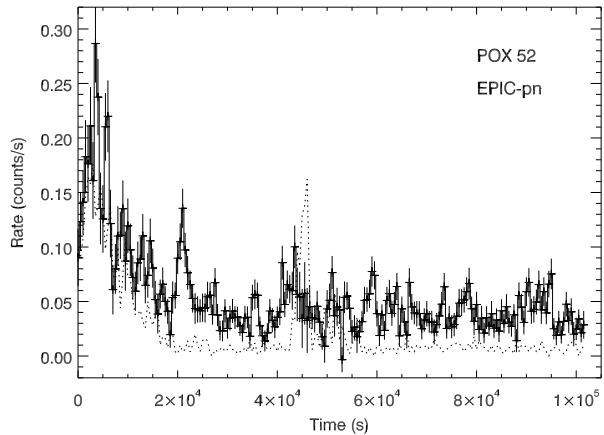}
  \caption{The background subtracted EPIC-pn lightcurve of Pox~52 with
    $500\s$ bins and in the energy band of $0.2-10\kev$. Also shown is
    the background level as dotted line in the same energy band.}
  \label{pox52_lc}
\end{figure}

\begin{figure*}
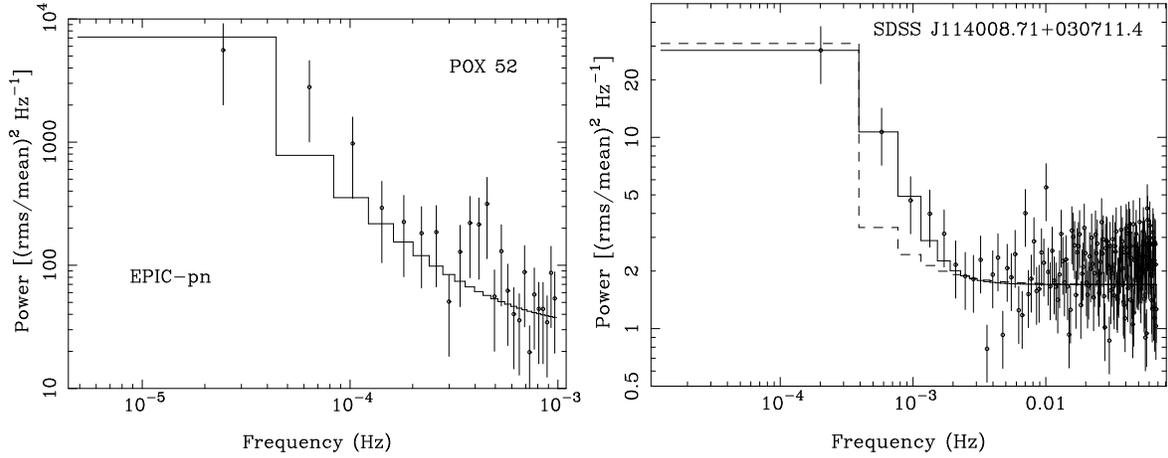

\centering \includegraphics[width=6cm,angle=-90]{fig4a.ps}
\includegraphics[width=6cm,angle=-90]{fig4b.ps}
\caption{{\it Left:} Power density spectrum of Pox~52 derived from the full band
  EPIC-pn lightcurve after excluding the first $15\ks$ exposure to
  minimized the effect of varying background. {\it Right:} Power
  density spectrum of J1140+0307 and the best-fit models -- a broken
  power law plus a constant (solid line) and a simple powerlaw plus a
  constant (dashed line).  }
\label{pds}
\end{figure*}

To derive the PDS, we re-extracted the source and background
lightcurves for POX~52 with time bins of $500\s$. Figure~\ref{f1}
shows the background corrected EPIC-pn lightcurve of POX~52.  Fig.~\ref{pox52_lc} also shows the EPIC-pn background
level relative to the net source level. The background rate varied in
the first $15\ks$ of observation and then it was steady except for the
a flare near the elapsed time of $\sim 45000\s$.  We calculated a
power density spectrum (PDS) using the background corrected EPIC-pn
light curves sampled at $500\s$. For this purpose, we excluded the
first $15\ks$ exposure to minimize the effect of varying
background. We used the `powspec' program in XRONOS. We rebinned the
PDS in logarithmic space with a binsize of 4 and performed the fitting
within the ISIS (version 1.4.8) spectral fitting environment. 
%We
%converted the PDS to equivalent energy spectra (power $\rightarrow$
%counts/bin; $\hz \rightarrow \kev$) using the {\tt sitar} timing
%package.  
The PDS is shown in Figure~\ref{pds} ({\it left panel}).  The power
arising from the Poisson errors has not been subtracted.  The Poisson
errors dominate the PDS above a frequency of $\sim
4\times10^{-4}\hz$. A simple power-law  (Power $\propto
f^{-\alpha})$ and a constant model resulted in a minimum $\chi^2 =
15.6$ for $23$ degrees of freedom (dof) with $\alpha = 1.6\pm0.3$. The constant
power was fixed to the value expected from the noise. There is a weak
excess of power at a frequncy of $\sim 3\times10^{-4}\hz$ that is
reminiscent of quasi-periodic oscillations (QPO). However, niether the
QPO-like feature nor a break in the PDS is statistically
significant. Hence we conclude that the PDS of POX~52 is consistent with a
simple power law and the determination of a break or the QPO-like
feature will require future high signal-to-noise long X-ray
observations. The non-detection of a break may be due to the dominance
of the Poission noise and the QPO-like feature above a frequency of
$\sim 2\times10^{-4}\hz$ below which the high frequency break is
likely to be present.

%Unless otherwise specified, the errors here and
%below are quoted at $90\%$ confidence level. 
%We then replaced the power-law with a broken power-law model. This
%model resulted in the minimum $\chi^2 = 93.5$ for $68$ dof. The
%best-fit parameters are listed in Table~\ref{t4}. The broken power-law
%model provided an improved fit only at a statistical significance
%level of $97.2\%$ compared to the power-law model according to the
%f-test. Thus we conclude that the PDS is suggestive of a break at
%$4.3_{-3.0}^{+1.8} \times 10^{-4}\hz$ with power-law indices
%$0.6_{-0.5}^{+0.2}$ below the break and $>2.3$ above the break.  The
%constant power due to Poisson noise is $27.9{\rm~(rms/mean)^2
%  Hz^{-1}}$.  The best-fit broken power-law plus constant model a
%plotted in Figure~\ref{pds}.

J1140+0307 is another AGN in our sample that has good signal-to-noise,
$\sim 40\ks$ long X-ray lightcurve. We have derived the PDS of
J1140+0307 using the EPIC-pn data. First we extracted a lightcurve
with $7.34\s$ bins. The source lightcurve was corrected for background
contribution though the background rate was stable throughout the
observation. As before we calculated the PDS and rebinned
logarithmically into bins of sizes 15. A simple power-law plus
constant model provided a minimum $\chi^2 = 213$ for $177$ dof with
power-law index steeper than $1.26$. Using a broken power-law model in
place of the simple power-law improved the fit. However, all the
parameters were not well constrained. We needed to fix the power-law
index $\alpha_2$ after the break at its best-fit value to calculate
the errors.  This procedure resulted in $\chi^2/dof = 203.6/176$. The
break frequency is in the range $(0.2-1.9)\times10^{-3}\hz$, index
$\alpha_1 = 0.1-1.8$, while the $\alpha_2$ was fixed at
$3$. Figure~\ref{pds} ({\it right panel}) compares the PDS, the
power-law model (dashed line) and the broken power-law model (solid
line). Table~\ref{t4} lists the PDS fit parameters. The broken
power-law model is an improved fit over the simple power-law model at
a confidence level of $99.5\%$, thus the departure from the simple
power-law is likely real. However, a long \xmm{} observation is
required to measure the break frequency reliably.

\begin{table}
%{\footnotesize
\centering
  \caption{Best-fit parameters for the power density spectrum of POX~52 \& J1140+0307 \label{t4}}
\begin{tabular}{llll} \tableline\tableline
 Model     &  Parameter &  POX~52 & J1140+0307 \\ \tableline
Power law  &  $\alpha$   &  $1.6\pm0.3$ & $>1.26$ \\
%          &  $n_{PDS}$    & $>0.0026$  & $1.7\times10^{-5}$ (not constrained) \\  
Constant   & $C$         & $30.0$ (fixed) & $1.7\pm0.1$  \\     
           & $\chi^2/dof$  & $15.6/23$ & $213/177$ \\  \tableline
Broken power-law & $\alpha_1$ & -- &  $1.0_{-0.9}^{+0.8}$ \\
           & $\alpha_2$         & --  & $3$ (fixed) \\
           & $f_{br}$($\hz)$   & -- & $0.7^{+1.2}_{-0.5}\times 10^{-3}$  \\
%          & $n_{PDS}$           & $>0.03$  & $0.00397_{-0.00396}^{+3.01}$  \\ 
Constant   &  $C$       & --  & $1.70_{-0.08}^{+0.07}$ \\       
          &  $\chi^2/dof$  & --  & $203.6/176$ \\ \tableline
\end{tabular}
\end{table}

\section{Spectral analysis}
%We extracted EPIC-pn spectra using a $25-40\arcsec$ circular regions
%with centroid at the source positions. We also extracted the
%background spectra using appropriate nearby circular regions free of
%sources.  We created appropriate response files using the SAS tasks
%{\tt rmfgen} and {\tt arfgen}. 
We have performed spectral analysis of the eight AGNs with IMBH. We
present the results based on  the EPIC-pn data only as these data have
the highest signal-to-noise. The spectra were analyzed with the
Interactive Spectral Interpretation System ({\tt ISIS, version
  1.4.8}).  The errors on the best-fit spectral parameters
  are quoted at  $90\%$ confidence level.
\subsection{NGC~4395 \label{n4395_spectral_analysis}}
\begin{figure*}
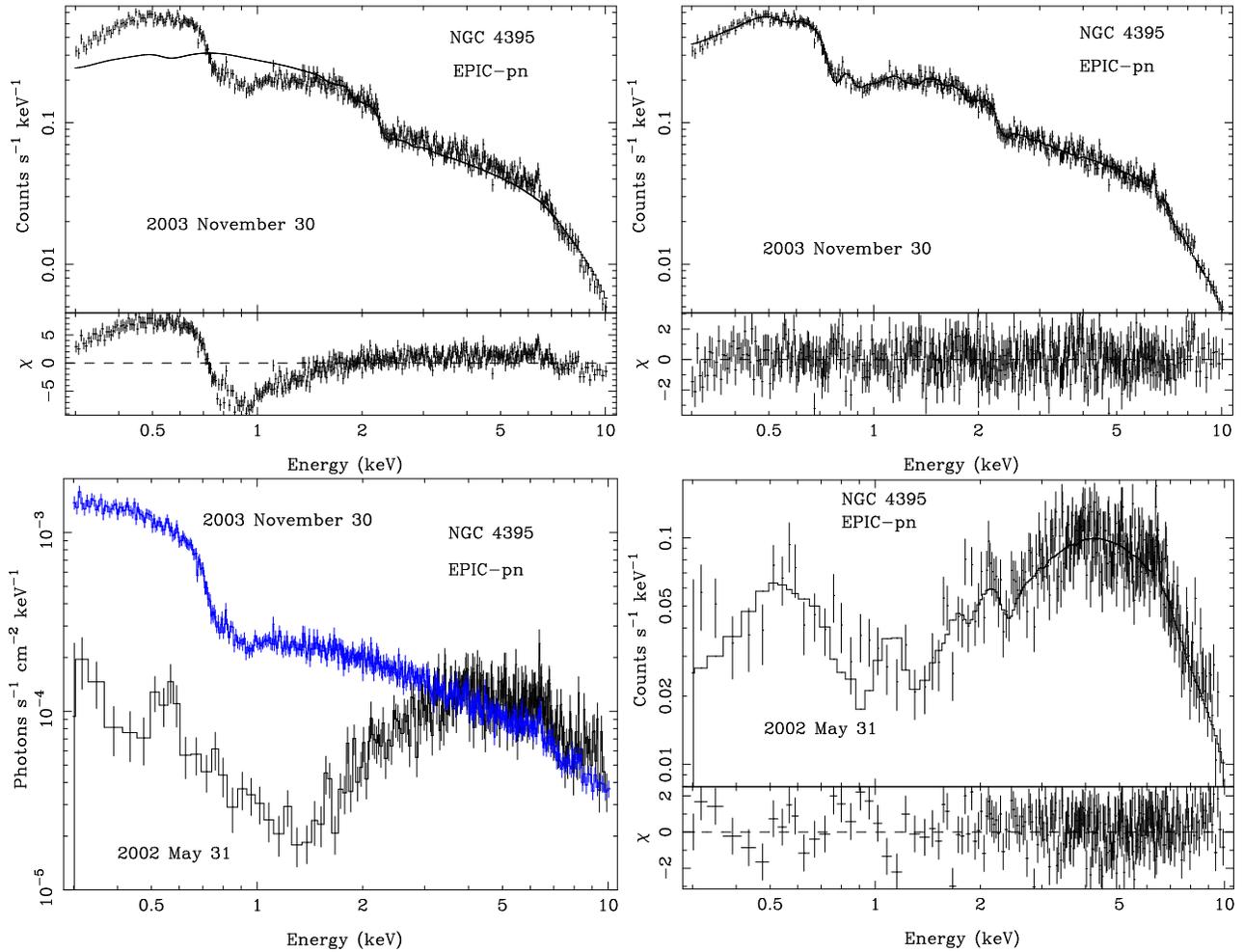

\centering
\includegraphics[width=6.5cm,angle=-90]{fig5a.ps}
\includegraphics[width=6.5cm,angle=-90]{fig5b.ps}
\includegraphics[width=6.5cm,angle=-90]{fig5c.ps}
\includegraphics[width=6.5cm,angle=-90]{fig5d.ps}
\caption{Results of the spectral analysis of NGC~4395. {\it Upper left:}
  EPIC-pn data obtained in 2003 November 30, the absorbed power-law
  model and the deviations. {\it Upper right:} EPIC-pn data of 2003
  November 30, the best-fit model consisting of a power-law and an
  iron K$\alpha$ line, modified by the Galactic absorption, intrinsic
  neutral and warm absorbers. {\it Lower left:} A comparison of
  unfolded EPIC-pn spectra of NGC~4395 obtained on 2003 November 30
  and 2002 May 31. {\it Lower right:} EPIC-pn data of 2002 May 31, the
  best-fit model and the deviations. }
\label{n4395_spectra}
\end{figure*}

We begin our spectral analysis with the high signal-to-noise spectrum
of NGC~4395 obtained on 2003 November 30. We grouped the EPIC-pn data
to a minimum of $100$ counts per spectral channel and fitted a simple
power-law (PL) model absorbed by the Galactic column to the $0.3-10\kev$
spectrum. This model provided unacceptable fit ($\chi^2/dof =
5627.7/485$). The observed data, the absorbed PL model and the
deviations are shown in the upper left panel of
Figure~\ref{n4395_spectra}. Evidently the X-ray emission from NGC~4395
is complex, showing evidence for emission/absorption features below
$2\kev$ and narrow iron K$\alpha$ line at $\sim 6.4\kev$. The lack
of emission near $1\kev$ and slight curvature in the $1-10\kev$ band
clearly suggest the presence of neutral/warm absorber covering the primary
source of X-ray emission fully or partially.

%As a next step, we added an accretion disk blackbody component ({\tt
%  diskbb} in XSPEC) to the power-law model to account for the soft
%X-ray excess emission. The diskbb+PL model improved the fit but it is
%not acceptable ($chi^2/dof = 1987.9/715$). The residuals showed broad
%emission/absorption features below $1\kev$, narrow iron line and
%curved spectrum above $7\kev$. Fitting a simple absorbed power-law
%plus Gaussian line model to the $2-10\kev$ provided a reasonably good
%fit ($\chi^2/dof = 425.7/388$), though slight curvature above $7\kev$
%is still present. The $2-10\kev$ power-law is flat with $\Gamma_X =
%1.20\pm0.04$ and modified by heavy ($N_H =
%9.1_{-1.6}^{+2.3}\times10^{21}{\rm~cm^{-2}}$. The heavy obscuration
%inferred from the $2-10\kev$ spectrum, the presence of soft excess
%emission below $0.7\kev$ and slight curvature above $7\kev$ suggest
%the presence of partially covering absorber. 
As a next step, we added a partially
covering neutral absorber component ({\tt zpcfabs}) to the PL
model and performed the fitting to the $0.3-10\kev$ data. The
addition of the partial covering absorber improved the fit
($\chi^2/dof = 2812.2/483$), though it is still not a good fit. The
covering fraction and absorption column suggested by the zpcfabs
component are $\sim 0.8$ and $\time 2.9\times10^{22}{\rm~cm^{-2}}$,
respectively. 
%The absorption column implied by the fully covering
%absorber ({\tt wabs}) is $1.6\times10^{21}{\rm~cm^{-2}}$, which is
%higher than the Galactic column ($N_H =
%1.85\times10^{20}{\rm~cm^{-2}}$). This suggests excess absorption,
%likely by the interstellar medium of NGC~4395.  Therefore, we fixed
%this column at the Galactic value ($N_H =
%1.85\times10^{20}{\rm~cm^{-2}}$) and added an additional neutral
%absorber, intrinsic to NGC~4395. This model, diskbb+PL modified by
%Galactic, intrinsic and partial covering absorption, resulted in
%$\chi^2/dof = 1663.2/713$. 
The poor quality of the fit is due mostly
to the broad emission features at $\sim 0.5-0.7\kev$ and absorption
features at $\sim 0.8-1\kev$ range. These features arise due to the
absorption and emission by warm absorbers.

To investigate the possible presence of a warm absorber medium, we
have created a warm absorber model using the spectral simulation code
CLOUDY version 7.02.01 \citep[last described
by][]{1998PASP..110..761F}. We used the ionizing continuum table
power-law available within CLOUDY ($f_\nu \propto \nu^{\alpha}$). We
assumed a spectral index of $\alpha = -0.9$ for the spectral range
between $10{\rm~ micron}$ and $50\kev$. The continuum has slopes
$f_\nu \propto \nu^{5/2}$ at low energies with an infrared break at
$10{\rm~microns}$ and $f_{\nu} \propto \nu^{-2}$ at high energies with
X-ray break at $50\kev$.  We also included the cosmic microwave
background radiation so that the incident continuum has nonzero
intensity at very long wavelengths. We assumed a plane parallel
geometry and calculated grids of models by varying the ionization
parameter and the total hydrogen column density. We also included UTA
features from \cite{2006ApJ...641.1227G} in our calculation. The grid
of models were imported to ISIS in the form of an XSPEC-style
multiplicative table model as described in \cite{2006PASP..118..920P}.

\begin{table*}
\centering
{\footnotesize
  \caption{Parameters for NGC~4395 and POX~52 obtained from the best-fitting models to the EPIC-pn data. \label{t5}}
  \begin{tabular}{llllll} \tableline\tableline
    Component &  Parameter\tablenotemark{1} & \multicolumn{3}{c}{NGC~4395}  & POX~52\\
              &            & 2003 November & 2003 January 3 & 2002 May
              31 & 2005 July 8 \\ \tableline
    Gal. Abs.   & $N_H$ ($10^{20}{\rm~cm^{-2}}$) & $1.85$(f) & $1.86$(f) & $1.86$(f)  & $3.85$(f) \\ \\
    Excess Abs. & $N_H$ ($10^{20}{\rm~cm^{-2}}$) & $5.2_{-0.2}^{+0.7}$
    & $5.2$(f) & $5.2$(f) & -- \\
    pcfabs    & $N_H$ ($10^{22}{\rm~cm^{-2}}$) & $36.4_{-3.1}^{+2.2}$ &
    $29.2_{-7.4}^{+13.0}$ & $8.0_{-1.2}^{+2.1}$  & $6.2\pm1.8$ \\
              & c.f. & $0.49_{-0.003}^{+0.0p}$  &
              $0.44_{-0.11}^{+0.08}$ & $0.78_{-0.15}^{+0.08}$ & $0.88_{-0.06}^{+0.04}$ \\
    WA(1)  &  $N_W$ ($10^{21}{\rm~cm^{-2}}$) & $5.8_{-0.3}^{+0.1}$ &
    $5.6_{-1.2}^{+1.0}$ & $9.5_{-2.1}^{+2.0}$  & $6.3_{-1.7}^{+1.8}$ \\
    & $log \xi$   &  $2.0_{-0.2}^{+0.1}$ & $2.2\pm0.1$ & $2.5\pm0.1$ & $1.8_{-0.8}^{+0.3}$\\
      WA(2) &  $N_W$  ($10^{21}{\rm~cm^{-2}}$) & $1.6_{-0.03}^{+0.3}$ &
    $2.5_{-0.5}^{+1.4}$ & $5.2_{-1.1}^{+1.8}$ & -- \\
     & $log \xi$ & $0.0_{-0p}^{+0.0001}$ &
     $0.001_{-0p}^{+0.4}$ & $0.009_{-0p}^{+0.4}$  & -- \\     
  WA(3) &  $N_W$  ($10^{20}{\rm~cm^{-2}}$) & $2.6_{-0.5}^{+0.1}$ &
    $1.8_{-0.8}^{+0.6}$ &  -- &  --  \\
    & $log \xi$   & $3.5_{-0.05}^{+0.03}$ & $3.7\pm0.3$ &  --  & --
    \\ \\  
MCD & $kT_{in}$($\ev$) & -- & -- & -- & $167_{-27}^{+37}$ \\
    & $n_{MCD}$\tablenotemark{2} & -- & -- & -- & $62.9_{41.1}^{+173.8}$ \\
   PL  & $\Gamma$    & $1.88_{-0.02}^{+0.01}$ & $1.88$(f) & $1.88$(f) & $2.0\pm0.3$ \\
    & $n_{PL}$\tablenotemark{3}     &
    $3.5_{-0.03}^{+0.02}\times10^{-3}$ & $3.5\times10^{-3}$(f) & $3.7_{-0.2}^{+0.3}\times10^{-3}$ & $1.7_{-0.7}^{+1.2}\times10^{-4}$
    \\ \\
  Gaussian & $E$ (kev) & $6.40\pm0.02$  & -- & $6.40$(f) & -- \\
                  & $\sigma$ (kev) & $<0.05$ & -- & $0.01$(f) & -- \\
                  & $f_{line}$\tablenotemark{4} & $(6.6\pm1.7)\times
                  10^{-6}$  & --  &$(1.4\pm0.5) \times10^{-5}$  & -- \\ \\
Flux     & $f_{0.3-2\kev}$\tablenotemark{5} & $9.4\times10^{-13}$ & $8.4\times10^{-13}$
& $1.0\times10^{-13}$  &  $4.4\times10^{-14}$  \\
         & $f_{2-10\kev}$\tablenotemark{5}  &  $5.5\times10^{-12}$   &
         $5.7\times10^{-12}$ & $6.0\times10^{-12}$   &  $2.7\times10^{-13}$ \\ \\
     & $\chi^2/dof$  & $508.3/473$  & $286.2/278$ & $323.6/260$ & $209.7/212$ \\ \tableline
\end{tabular}

\tablenotemark{1}{`(f)' indicates that the parameter value was
  fixed. `p' in the upper or lower error value indicates that the
  upper or lower confidence limit did not converge.}
\tablenotemark{2}{MCD normalization in units of $n_{MCD}=(R_{in}/{\rm
    km})/(D/10{\rm kpc})$, where $R_{in}$ is the inner radius and $D$ is
  the distance.}
\tablenotetext{3}{Power-law normalization in units of {$\rm photons~keV^{-1}~cm^{-2}~s^{-1}$} at $1\kev$.}
\tablenotetext{4}{Line flux in $\rm photons~keV^{-1}~cm^{-2}~s^{-1}$.}
\tablenotetext{5}{Observed flux in units of ${\rm ergs~cm^{-2}~s^{-1}}$.}
}
\end{table*}

The free parameters of the warm absorber model are absorption column
($N_{W}$), ionization parameter ($log \xi$, where $\xi = L/nr^2$) and
the redshift ($z$).  The use of the warm absorber model improved the
fit to $\chi^2/dof = 906.9/481$. The remaining features in the residuals
of the data and model below $2\kev$ are broad emission features at
$\sim 0.6\kev$, absorption features at $\sim0.7\kev$ and $\sim
0.9\kev$ and the lack of emission below $0.4\kev$. The features suggest
additional warm absorber components. Adding a second warm absorber
component further improved the fit to $\chi^2/dof =
854.4/479$. However the absorption feature at $\sim 0.7\kev$, likely
the absorption edge due to O~VII, is still not modeled. Adding a third
warm absorber component to account for the $\sim 0.7\kev$ absorption
feature improved the fit to $\chi^2/dof = 754.2/477$. Examination of
the residuals shows a lack of emission below $\sim 0.4\kev$ suggesting
additional neutral absorption. We have kept fixed the neutral column
at the Galactic value in all the above fits. Adding a neutral absorber
component in addition to the Galactic column improved the fit to
$\chi^2/dof = 548.3/476$. Finally, addition of a narrow Gaussian line
for the iron K$\alpha$ line provided an excellant fit
($\chi^2/dof=508.3/473$). The best-fit photon index is $\sim 1.88$.
We note that the $5-10\kev$ spectrum is consistent with an absorber
power-law with a photon index $\Gamma \sim 1.84$. Evidently the
multiple warm and neutral absorbers strongly modify the spectrum below
$\sim 5\kev$. The parameters of the best-fit model are listed in
Table~\ref{t5} and the observed EPIC-pn spectrum of 2003 November 30, the
best-fit model and the deviations are shown in the upper right panel
of Fig.~\ref{n4395_spectra}.

We have also analyzed the EPIC-pn spectrum of NGC~4395 obtained on
2002 May 31 with a net exposure of $12.5\ks$. The EPIC-pn data were
grouped to a minimum of $20$ counts per spectral
channel. Fig.~\ref{n4395_spectra} ({\it lower left panel}) shows a
comparison of the unfolded EPIC-pn spectra of NGC~4395 obtained on
2002 May 31 and 2003 November 30 in a model independent way. Unlike
XSPEC, the unfolded spectrum in ISIS is derived in a model-independent
way as follows:
\begin{equation}
f_{unfold} (I) = \frac{[C(I) - B(I)]/\Delta t}{\int {R(I,E)A(E)dE}},
\end{equation}
where C(I) is the number of total counts in the energy bin $I$, $B(I)$
is the number of background counts, $\Delta t$ is exposure time,
$R(I,E)$ is the normalized response matrix and $A(E)$ is the effective
area at energy $E$.  This definition produces a spectrum that is
independent of fitted model. It is clear from
Figure~\ref{n4395_spectra} that the X-ray spectrum of NGC~4395 varied
drastically below $\sim 3\kev$ but remained similar 
at harder energies. The presence of the dip near $1\kev$ in the
spectrum of 2002 May clearly suggests heavy obscuration by neutral or
partially ionized material. Thus we expect that the best-fit model
inferred from 2003 November observation should also be appropriate for
2002 May observation except for different absorption columns of either
neutral or warm absorber components. Thus we fit the model of 2003
November observation to the EPIC-pn data obtained on 2002 May with the
parameters fixed at their best-fit values and we varied the parameters
of the absorbing components. First we varied the parameters of the
partial covering absorption model and then those of the warm absorber
components as required to obtain a good fit. We additionally needed to
vary the normalizations of the power law and the narrow Gaussian line
at $6.4\kev$. The warm absorber model with the highest ionization
parameter for the 2003 November data was not required for the 2002 May
data. The fully covering intrinsic neutral absorption did not
vary. The best-fit parameters for the 2002 May data are listed in
Table~\ref{t5} and the observed data, the best-fit model and the
deviations are shown in the lower right panel of
Fig.~\ref{n4395_spectra}. As expected, the covering fraction was
higher ($\sim 80\%$) in 2002 May, compared to $\sim50\%$ in 2003
November.  As seen in Table~\ref{t5}, most of the spectral variability
is caused by the change in the partial covering and warm absorbers.

The model for the 2003 November data also describes the EPIC-pn data
obtained on 2003 Jan 3 except for small changes in the parameters of the
partial covering and warm absorber components (see Table~\ref{t5}). An
iron K$\alpha$ line was not detected in the observation of 2003 Jan
3. The model provided $\chi^2/dof = 286.2/278$ with the partially covering
absorption model parameters
$N_H = 2.9_{-0.7}^{+1.3}\times10^{23}{\rm~cm^{-2}}$ and covering
fraction of $44_{11}^{+8}\%$. Thus the covering fraction was
$\sim80\%$ in 2002 May, $\sim50\%$ in 2003 Jan and 2003 November.

The best-fit model for the 2003 November data suggests absence of soft
X-ray excess emission below $2\kev$ that is observed from many NLS1
galaxies. We have estimated an upper-limit to the strength of the soft X-ray
excess emission. First we fitted a simple absorbed power-law model to
the $5-10\kev$ data in order to minimize the effect of multiple
absorber components on the intrinsic spectral shape and obtained
$\Gamma_{5-10\kev} = 1.84$. To estimate the upper-limit to the soft excess
emission, we added a multicolor disk blackbody component (MCD) to the
best-fit model to the 2003 November data and fixed the photon index to
$\Gamma=\Gamma_{5-10\kev}$ and calculated the upper-limit to the flux
of the MCD component.  We obtained an upper-limit of $22\%$ for the
strength of the soft excess emission relative to the PL
component in the $0.3-2\kev$ band.

\subsection{POX~52}
\begin{figure*}
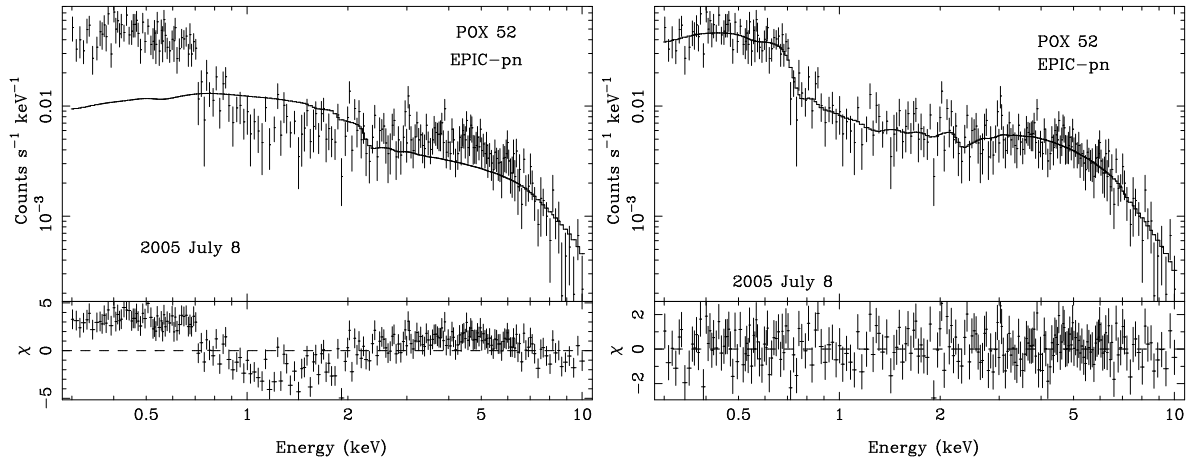

\centering \includegraphics[width=6cm,angle=-90]{fig6a.ps}
 \includegraphics[width=6cm,angle=-90]{fig6b.ps}
 \caption{{\it Left:} Observed EPIC-pn spectrum, absorbed power-law
   model and the deviations. {\it Right:} Observed EPIC-pn spectrum,
   the best-fit model consisting of a power law and an MCD modified by
   a multiple absorbing components -- the Galactic, partially covering
   neutral absorption and a warm absorber. Also shown are the
   deviations of the observed data from the best-fit model.}
\label{pox52_spectra}
 \end{figure*}

The EPIC-pn spectrum of POX~52 was grouped to a minimum of $20$ counts
per spectral channel.  First we fitted a simple absorbed 
PL model to the pn data in the $0.3-10\kev$ band.  This model
provided an unacceptable fit ($\chi^2/dof = 1004.5/217$). We have
plotted the observed data, the power-law model and the $\chi$
residuals in Figure~\ref{pox52_spectra} ({\it left panel}). The
deviations of the data from the PL model clearly show a soft X-ray
excess emission below $0.7\kev$ and a broad hump in the $2.5-7\kev$
band. The deviations are qualitatively similar to that for NGC~4395
albeit with reduced strength of soft excess and no iron K$\alpha$ line
(see Fig.~\ref{n4395_spectra}, upper left panel).  Adding a partial
covering neutral absorber to the PL model improved the fit
($\chi^2/dof = 429.8/216$) but the fit is not acceptable. The
residuals showed absorption features at energies $\sim 0.7-1.1\kev$,
suggestive of the presence of warm absorber medium. We used the same
warm absorber model created earlier for NGC~4395 to fit the EPIC-pn
spectrum of POX~52 (see Section~\ref{n4395_spectral_analysis}). The
addition of the warm absorber model to the PL model modified by a
partial covering neutral absorber improved the fit, providing an
excellent fit ($\chi^2/dof = 212.2/214$). The best-fit photon index is
steep ($\Gamma = 2.4\pm0.2$). However, a simple absorbed power-law
model fit to the $5-10\kev$ band data provided $\Gamma =
2.0_{-0.4}^{+0.6}$. The discrepancy is most likely due to the presence
of multiple absorbers as it is difficult to infer the intrinsic
spectral shape if most of the observed band is strongly affected by
absorbers. Therefore, we fixed $\Gamma = 2.0$ in our best-fit model to
the full band spectrum and performed the fitting. The fit worsened to
$\chi^2/dof = 230.5/215$. The residuals suggested possible presence
for a soft X-ray excess component. Additon of a multicolor disk
blackbody (MCD) model and varying the photon index improved the fit to
$\chi^2/dof = 209.7/212$. The best-fit photon index is $\Gamma =
2.0\pm0.3$ similar to that obtained for the $5-10\kev$ band data. The EPIC-pn data, the
best-fit model and the deviations are plotted in
Fig.~\ref{pox52_spectra} (see right panel). The best-fit parameters
are listed in Table~\ref{t5}.
In the $0.3-2\kev$ band, the soft X-ray excess and the power-law
component have similar fluxes.

\subsection{SDSS AGNs with IMBH}
There are seven observations of six SDSS AGNs with IMBHs that are
suitable for spectral fitting. All the EPIC-pn spectra were grouped to
a minimum of $20$ counts per spectral channel. These spectra are
relatively simple compared to that of NGC~4395 or POX~52. As a first
step, we fitted a simple power-law modified with Galactic absorption
to the individual spectra.
The residuals of the
observed data from the power-law model showed evidence for weak soft
excess emission in the case of J0249-0815, J1140+0307 and
J1357+6525. Addition of a multicolor disk blackbody component to the
PL model improved the fit marginally for J0249-0815 only at a level of
$97.4\%$ ($\Delta \chi^2 = -8.5$ for two additional parameters) based
on the f-test. The best-fit parameters are $kT_{in} =
169_{-50}^{+43}\ev$ and $\Gamma = 1.72\pm0.30$. For J1140+0307, the
addition of the MCD component provided significant improvement over
the power-law model ($\Delta\chi^2 = -64.6$ for one additional
parameter). Varying the absorption column did not improve the fit
further. The best-fit parameters are $kT_{in} = 181_{-7}^{+3}\ev$ and
$\Gamma = 2.4\pm0.1$.  J1357+6525 has two observations. Both data show
evidence for weak soft excess emission. The spectral shapes are
similar withing errors for both the observations except for a narrow
iron K$\alpha$ line at $6.65\kev$ which has been detected in the first
observation at a level of $99.3\%$ based on the F-test. The best-fit
parameters for the SDSS AGNs are listed in Table~\ref{t6}.

\begin{figure*}
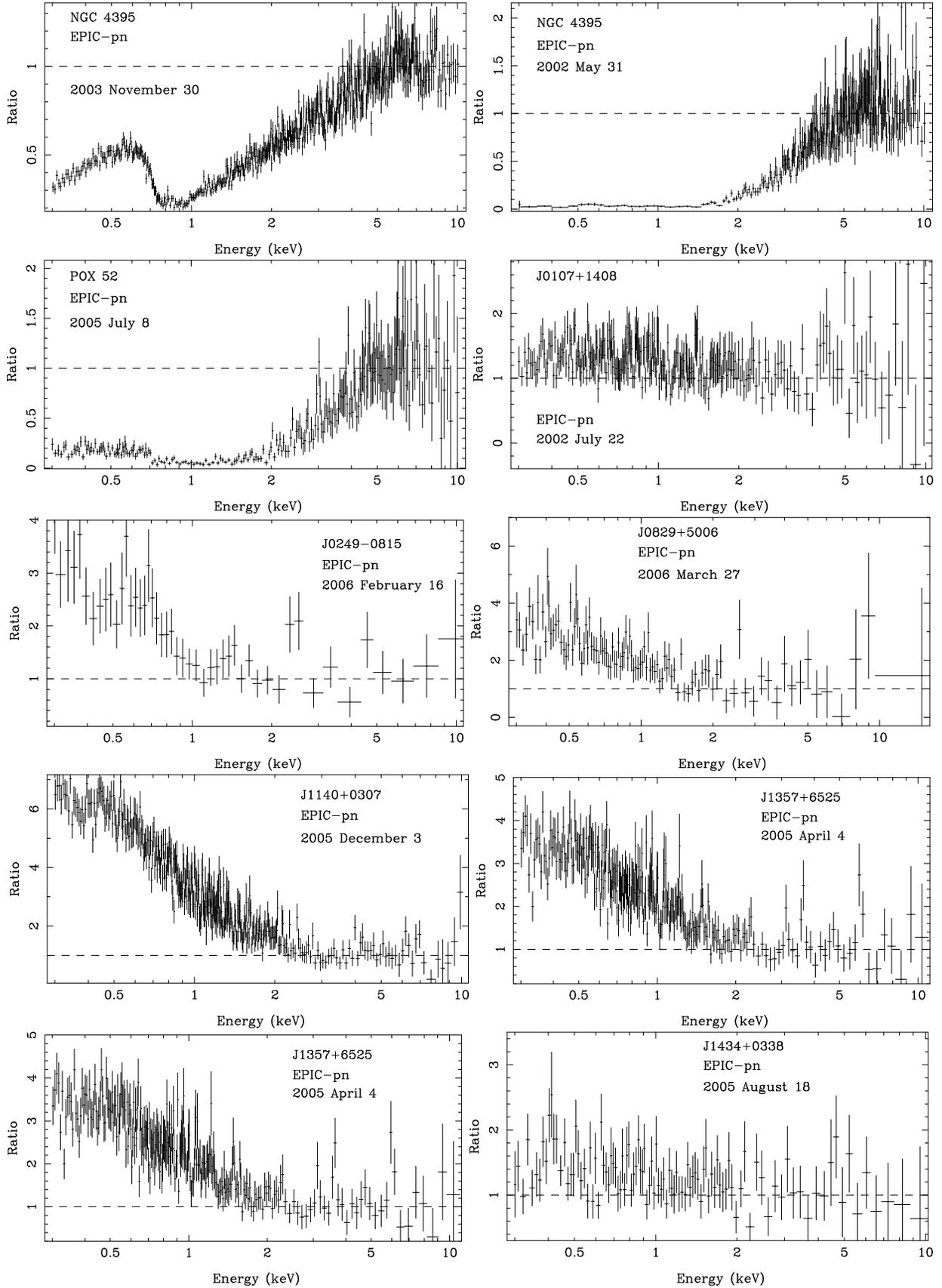

\centering
\includegraphics[width=4.5cm,angle=-90]{fig7a.ps}
\includegraphics[width=4.5cm,angle=-90]{fig7b.ps}
\includegraphics[width=4.5cm,angle=-90]{fig7c.ps}
\includegraphics[width=4.5cm,angle=-90]{fig7d.ps}
\includegraphics[width=4.5cm,angle=-90]{fig7e.ps}
\includegraphics[width=4.5cm,angle=-90]{fig7f.ps}
\includegraphics[width=4.5cm,angle=-90]{fig7g.ps}
\includegraphics[width=4.5cm,angle=-90]{fig7h.ps}
\includegraphics[width=4.5cm,angle=-90]{fig7i.ps}
\includegraphics[width=4.5cm,angle=-90]{fig7j.ps}
\caption{The ratios of EPIC-pn data and the best-fitting power-law
  model modified by the line of sight Galactic absorption. The power
  law was fitted to $2.5-10\kev$ data for the IMBH AGNs discovered
  with the SDSS and extrapolated to lower energies to calculate the
  ratios. For NGC~4395 and POX~52, the ratios were obtained by fitting
  the absorbed power law to the $5-10\kev$ band as the primary
  continua of the two AGNs are strongly modified below $5\kev$ by the
  multiple absorbing components.}
\label{sdss_agn_spectra}
\end{figure*}

The simple power-law model provided statistically acceptable fits to
the pn data of J0107+1408, J0829+5006 and J1434+0338 with photon
indices $\Gamma = 2.4\pm0.1$, $2.65\pm0.08$, $2.05\pm0.07$,
respectively. The steep photon indices suggest possible presence of
the soft X-ray excess emission from these AGNs. However, addition of
an MCD component did not improve any of the fits statistically at
$95\%$ level. To further investigate the presence of the soft excess
component, we fitted the absorbed power-law model only to the hard
$2.5-10\kev$ and found generally flatter indices.  To estimate the
strength of the soft excess emission from these AGNs, we added MCD
components to the power-law model and fixed the power-law index to the
best-fit values obtained for the $2.5-10\kev$ data. The fit qualities
are similar to the simple power-law fits. Thus the data are consistent
with the presence of the soft excess but do not require this component
statistically. Therefore, we calculated $90\%$ upper-limits to the
strength of the soft excess emission. The upper limits are listed in
Table~\ref{t7}.  In Figure~\ref{sdss_agn_spectra}, we show the
strength of the soft X-ray excess from the SDSS AGNs with IMBHs as the
ratios of the observed data and the best-fit $2.5-10\kev$ power-law
model extrapolated to lower energies.  For comparison, we have also
plotted the ratios for NGC~4395 and POX based on the best-fit absorbed
power law to the $5-10\kev$ data as the X-ray emission  below $\sim
5\kev$ is strongly modified by the intrinsic neutral and warm
absorbes.

\begin{table*}
\centering
  \caption{The best-fitting  spectral
    parameters for SDSS AGNs. \label{t6}}
\begin{tabular}{lccccccccr}   \tableline\tableline
Object & Data & Model & Cold $N_H$ & $kT_{in}$  & $\Gamma$ &
($E_{FeK\alpha}$, $\sigma$, $f_{FeK\alpha}$\tablenotemark{1}) & $f_{soft}$\tablenotemark{2} & $f_{hard}$\tablenotemark{2} & $\chi^2/dof$ \\ 
       &             &       &  ($10^{20}{\rm~cm^{-2}}$) & & &
       ($\kev$), ($\kev$),  &
        &  &     \\ \tableline
J0107+1408  & (1) & PL & $4.5\pm1.0$ & -- & $2.4\pm0.1$ & --
& $3.4$ & $1.9$ & $213.1/236$ \\
            &    & PL+MCD & $5.1_{-1.6}^{+2.0}$ & $157_{-28}^{+49}$ & $2.20$(f) & -- &  $3.3$ & $2.2$ & $209.9/235$ \\  
J0249--0815 & (1) & PL & $3.67$(f) & -- &
$2.2_{-0.1}^{+0.2}$ &-- & $3.4$ & $2.5$ &
$52.4/46$ \\
                     & (1)    & MCD+PL & $3.67$(f) & $169_{-41}^{+33}$ &
                     $1.6\pm0.3$ &-- & $3.2$ &
                     $4.1$ & $37.9/44$ \\ 
J0829+5006 & (3) & PL&  $4.08$(f) & -- & $2.64\pm0.08$ &-- &
$12.2$ & $4.5$ & $102.4/101$ \\
          &      & PL+MCD &$4.08$(f) & $181_{-22}^{+23}$ & $2.04$(f) &
          -- & $11.8$ & $6.6$ & $100.6/100$ \\  
J1140+0307 & (1) & PL & $3.9\pm0.6$ & -- & $2.93\pm0.06$ &-- & $7.8$ & $1.8$ & $424.2/333$ \\ 
                   & (1) & MCD+PL & $1.91$(f) & $181_{-7}^{+8}$ &
$2.40_{-0.09}^{+0.08}$ &-- & $7.7$ & $2.4$ &
$359.6/332$ \\
J1357+6525 & (1) & PL & 
$2.4_{-0.85}^{+0.90}$ & -- & $2.60\pm0.08$ &-- & $5.4$ &
$1.9$ & $287.5/237$ \\
                    & (1) & MCD+PL & $1.36$(f) & $184\pm16$ & $2.18\pm0.13$ &
                    -- &  $5.3$ & $2.5$ & $266.8/236$ \\
                    & (1) & MCD+PL+GL & $1.36$(f) & $186_{-17}^{+18}$
                    & $2.23\pm0.13$ & $6.65_{-0.10}^{+0.05}$,
                    $0.01$(f), $1.7\pm0.9$ & $5.4$ & $2.5$ &
                    $255.6/234$ \\
                    & (3) & PL & $2.1_{-1.0}^{+1.1}$ & -- &
                    $2.5\pm0.1$ & -- & $5.0$ & $2.2$ &
                    $207.7/160$ \\
                    & (3) & MCD+PL & $1.36$(f) & $186_{-17}^{+16}$ &
                    $1.93\pm0.15$ & -- &  $4.9$ & $3.0$ &
                    $176.8/160$ \\ 
J1434+0338 & (1) & PL & $2.43$(f) & -- & $2.05\pm0.07$ & -- & $1.3$ &
$1.2$ & $101.0/103$ \\
           &     & PL+MCD & $2.43$(f) & $208_{-80}^{+89}$ & $1.86$(f) &
           -- & $1.3$ & $1.3$ & $101.3/102$ \\
\tableline
\end{tabular}
\tablenotetext{1}{Iron line flux in units of $10^{-6}{\rm~photons~cm^{-2}~s^{-1}}$.}
\tablenotetext{2}{$f_{soft}$ and $f_{hard}$ are flux in the soft
  ($0.3-2\kev$) and hard ($2-10\kev$) bands in units of $10^{-13}{\rm
         ergs~cm^{-2}~s^{-1}}$.}
\end{table*}

\begin{table*}
\centering
\caption{Strength of soft X-ray excess emission, luminosity \& accretion rates for AGNs with IMBHs. \label{t7}}
\begin{tabular}{lccccc} \tableline\tableline
  Object & $\frac{L_{MCD}}{L_{PL}}$ & $\frac{L_{MCD}}{L_{PL}}$  &  $L(0.3-10\kev)$ & $L(2-10\kev)$ & $\frac{L_X}{L_{Edd}}$ \\ 
   & ($0.3-2\kev$) & ($0.6-10\kev$) & ($\rm ergs~s^{-1}$) & ($\rm ergs~s^{-1}$) &    \\ \tableline
  NGC~4395      & $<0.22$ & $<0.02$ & $4.5\times10^{40}$ & $2.3\times10^{40}$ & $1.0\times10^{-3}$\\
  POX~52        & $1.0$  & $0.24$  & $1.4\times10^{42}$ & $4.1\times10^{41}$ & $0.07$ \\
  J0107+1408    & $<0.28$  & $<0.08$  & $9.2\times10^{42}$ & $2.7\times10^{42}$ & $0.11$ \\
  J0249--0815   & $0.80$ & $0.11$ & $1.6\times10^{42}$ & $8.4\times10^{41}$ & $0.11$ \\
  J0829+5006    & $<1.6$  & $<0.4$  & $9.1\times10^{42}$ & $1.9\times10^{42}$ & $0.14$  \\
  J1140+0307    & $0.61$ & $0.24$ & $1.8\times10^{43}$ & $3.8\times10^{42}$ & $0.24$ \\
  J1357+6525    & $0.42$ & $0.14$ & $2.3\times10^{43}$ & $6.6\times10^{42}$ & $0.19$ \\
  J1434+0338    & $<1.9$  & $<0.5$  & $4.8\times10^{41}$ & $2.1\times10^{41}$ & $0.02$ \\  \tableline
\end{tabular}
\end{table*}

\section{UV emission and optical-to-X-ray spectral index}
An additional advantage of \xmm{} is the availability of the OM that
provides optical/UV data simultaneously with the X-ray
observations. All AGNs with IMBHs were observed with the OM in the
imaging mode with one or more optical/UV filters. The AGNs were
observed in the UV for the first time, therefore we provide here
standard UV magnitudes and fluxes based on the OM observations. The OM
data were processed using the SAS task {\tt omichain}. In all cases,
multiple exposures were taken in the same filter and the standard
magnitudes and flux densities were derived based on the average count
rates (Cohen 2004). We list the average count rates, standard
magnitude and flux densities for each of the observations in
Table~\ref{tab8}. We corrected the flux densities for the Galactic
extinction by adopting the extinction law in
\cite{1989ApJ...345..245C}. We calculated the color excesses E(B-V)
due to the Galactic reddening from the Galactic hydrogen column
density ($N_H$) along the lines of sight to individual AGNs using the
relation
\begin{equation}
  N_H = 5.8\times10^{21} \times E(B-V) {\rm~cm^{-2}}
\end{equation}
\citep{1978ApJ...224..132B}. The unreddened flux densities are
listed in Table~\ref{tab8}.

We have also calculated the optical-to-X-ray spectral index ($\alpha_{ox}$) defined as 
\begin{equation}
\alpha_{ox} = -0.386 log\left(\frac{L_{\nu}(2500{\rm \AA})}{L_{\nu}(2\kev)}\right)
\end{equation}
where $L_{\nu}$($2500{\rm \AA})$ and $L_{\nu}$($2\kev$) are the
intrinsic luminosity densities at $2500{\rm \AA}$ and ${\rm 2keV}$,
respectively. We have derived the flux densities from the OM and
EPIC-pn observations. NGC~4395 was observed in four filters UVW1,
UVW2, U and B during the third observation, we calculated the flux
density at $2500{\rm \AA}$ by fitting a power-law $f_{\nu} \propto
\nu^{\alpha_{UV}}$ to the unreddened OM flux densities. For other AGNs
with flux densities available in two OM filters, we calculated
$f_{\lambda}(2500{\rm \AA})$ by directly calculating $\alpha_{UV}$ and
the normalization of the power law. For POX~52, J0249--0815 and
J1434+0338 with flux densities available in single UV filter, we
assumed a UV spectral index, $\alpha_{UV} = -0.9$, the same as that
derived for NGC~4395. The UV spectral indices and the derived
monochromatic fluxes at $2500{\rm \AA}$ are listed in
table~\ref{tab9}. The flux densitities at $2\kev$ were derived from the
best-fit models to the EPIC-pn data obtained simultaneously with the
OM data and were corrected for the Galactic and any neutral
or warm absorption. The flux
densities at $2\kev$ are also listed in Table~\ref{tab9} along with the
$\alpha_{ox}$ values.

\begin{table*}
\centering
\caption{\xmm{} OM flux measurements \label{tab8}}
\begin{tabular}{lccccccc} \tableline\tableline
  Object & Observation & Filter & Effective   & Count rate\tablenotemark{1} & Magnitude\tablenotemark{3} & \multicolumn{2}{c}{$f_{\lambda}$(${\rm ergs~cm^{-2}~s^{-1}~\AA^{-1}}$)}  \\
  &             &        & wavelength  & (${\rm counts~s^{-1}}$) &        & Reddened & Unreddened     \\ \tableline
  NGC~4395 & (2)  & UVW1 & $2905{\rm \AA}$ & $4.0\pm0.03$ & $17.06\pm0.01$ & $(1.9\pm0.02)\times10^{-15}$ & $(2.25\pm0.02)\times10^{-15}$  \\
  &      & U & $3472{\rm \AA}$ & $9.3\pm0.06$  &   $16.76\pm0.01$ & $(1.8\pm0.01)\times 10^{-15}$ & $(2.08\pm0.01)\times 10^{-15}$  \\
  & (3)  & UVW2 & $2070{\rm \AA}$ &  $0.30\pm0.007$ & $17.88\pm0.03$ & $(1.7\pm0.04) \times10^{-15}$ & $(2.24\pm0.05)\times10^{-15}$ \\
  &      & UVW1 &  $2905{\rm \AA}$ &$3.14\pm0.023$ & $17.32\pm0.01$  &$(1.5\pm0.01)\times10^{-15}$ & $(1.78\pm0.01)\times10^{-15}$ \\
  &      & U    & $3472{\rm \AA}$ &$7.75\pm0.036$ & $16.97\pm0.01$ & $(1.5\pm0.01) \times10^{-15}$ & $(1.73\pm0.01)\times10^{-15}$ \\ 
  &      & B    & $4334{\rm \AA}$ & $8.07\pm0.078$ & $16.81\pm0.01$ & $(1.0\pm0.01)\times10^{-15}$ & $1.13\pm0.01)\times10^{-15}$ \\ 
  & (4)  & UVW2 & $2070{\rm \AA}$ &$0.31\pm0.015$ &  $17.83\pm0.05$ &$(1.8\pm0.09)\times10^{-15}$ & $(2.38\pm0.11)\times10^{-15}$  \\ 
  &      & U    & $3472{\rm \AA}$ & $4.96\pm0.115$ & $17.45\pm0.03$ & $(9.6\pm0.22)\times 10^{-16}$ & ($1.11\pm0.25)\times10^{-15}$  \\ \\
  POX~52   & (1) & UVM2 & $2298{\rm \AA}$ & $0.41\pm0.004$ & $18.37\pm0.01$ & $(9.1\pm0.09)\times10^{-16}$ & $(1.56\pm0.01\times10^{-15}$   \\ 
  \\
  J010712+140844 & (1) &  UVM2 & $2298{\rm \AA}$ & $0.08\pm0.005$ & $20.10\pm0.07$ & $(1.8\pm0.11)\times10^{-16}$ & $(2.92\times0.18)\times10^{-16}$  \\ 
  &     &  UVW1 & $2905{\rm \AA}$ & $0.34\pm0.011$ & $19.72\pm0.03$ & $(1.7\pm0.054)\times 10^{-16}$ & $(2.33\pm0.07)\times 10^{-16}$  \\ 
  \\
  J024912.86--081525.6 & (1) & UVW1 & $2905{\rm \AA}$ & $0.37\pm0.023$ & $19.64\pm0.07$ & $(1.8\pm0.109)\times10^{-16}$ & $(2.52\pm0.15)\times10^{-16}$ \\
  \\ 
  J082912.67+500652.3 & (1) & UVW1 & $2905{\rm \AA}$ & $0.91\pm0.031$ & $18.67\pm0.04$  & $(4.4\pm0.15) \times10^{-16}$ & $(6.41\pm0.22)\times10^{-16}$ \\
  % & (2) & NU & \\
  & (3) & UVW1 & $2905{\rm \AA}$ & $0.93\pm0.026$ & $18.65\pm0.03$ & $(4.5\pm0.13)\times10^{-16}$ & $(6.55\pm0.20)\times10^{-16}$  \\
  \\
  J114008.71+030711.4 & (1) & UVM2 & $2298{\rm \AA}$ & $0.26\pm0.007$ & $18.87\pm0.03$ & $(5.77\pm0.15)\times10^{-16}$ & $(7.55\pm0.20)\times10^{-16}$   \\
  &     & UVW1 & $2905{\rm \AA}$ & $0.73\pm0.012$ & $18.90\pm0.02$ & $(3.53\pm0.01) \times10^{-16}$ & $(4.21\times0.02)\times10^{-16}$   \\
  \\ 
  J135724.52+652505.8 & (1) & UVM2 & $2298{\rm \AA}$ & $0.21\pm0.006$ & $19.08\pm0.03$ & $(4.7\pm0.13)\times10^{-16}$ & $(5.69\pm0.16)\times10^{-16}$  \\ 
  &     & UVW1 & $2905{\rm \AA}$ & $0.69\pm0.010$ & $18.97\pm0.02$ & $(3.3\pm0.01) \times10^{-16}$ & $(3.74\pm0.01)\times10^{-16}$ \\ 
  & (2) & UVM2 & $2298{\rm \AA}$ & $0.21\pm0.008$ & $19.11\pm0.04$ & $(4.6\pm0.18)\times10^{-16}$ & $(5.57\pm0.22)\times10^{-16}$ \\ 
  & (3) & UVM2 & $2298{\rm \AA}$ & $0.19\pm0.008$ & $19.20\pm0.04$ & $(4.2\pm0.18)\times10^{-16}$ & $(5.10\pm0.22)\times10^{-16}$  \\
  \\
  J143450.62+033842.5 & (1) & UVM2 & $2298{\rm \AA}$ & $0.18\pm0.009$ & $19.26\pm0.05$ & $(4.0\pm0.20)\times10^{-16}$ & $(5.63\pm0.28)\times10^{-16}$ \\ 
  \tableline 
\end{tabular}
\tablenotetext{1}{Corrected rate}
%\tablenotetext{2}{Fluxes are AB system and in the units of ${\rm ergs~cm^{-2}~s^{-1}~\AA^{-1}}$.}
\tablenotetext{3}{UVW1, UVW2, U and B are in AB magnitudes}
\end{table*}

\begin{table*}
\caption{$\alpha_{ox}$ for AGNs with IMBHs \label{tab9}}
\begin{tabular}{lccccccc} \tableline\tableline
Object    &    Observation & $\alpha_{UV}$ &  $f_{\nu}$($2500{\rm \AA}$) & $f_{\nu}$($2\kev$) &
$L_{\nu}$ ($2500{\rm \AA}$) & $L_{\nu}$($2\kev$) & $\alpha_{ox}$ \\
          &  & & (${\rm ergs~cm^{-2}~s^{-1}~Hz^{-1}}$) &
          (${\rm ergs~cm^{-2}~s^{-1}~Hz^{-1}}$) &  (${\rm
            ergs~s^{-1}~Hz^{-1}}$) & (${\rm
            ergs~s^{-1}~Hz^{-1}}$) &  \\  \tableline
NGC~4395   &   (3) & $-0.9$  & $4.1\times10^{-27}$  &  $1.3\times10^{-29}$ &
$8.8\times10^{24}$ &  $2.7\times10^{22}$ & $-0.97$ \\
POX~52     &  (1) & $-0.9$\tablenotemark{a}  & $3.0\times10^{-27}$  &  $5.7\times10^{-31}$ &
$2.9\times10^{27}$ & $5.5\times10^{23}$  &  $-1.44$ \\
J0107+1408 & (1) & $-1.0$  & $5.6\times10^{-28}$  &  $3.3\times10^{-31}$ &
$7.9\times10^{27}$ &  $4.7\times10^{24}$ &  $-1.24$ \\
J0249-0815 & (1)  & $-0.9$\tablenotemark{a} & $4.1\times10^{-28}$  &  $3.1\times10^{-31}$ &  $8.0\times10^{26}$ &  $6.0\times10^{23}$ &  $-1.20$ \\
J0829+5006 & (1)  & $-0.9$\tablenotemark{a} & $2.1\times10^{-27}$  &  $9.2\times10^{-31}$ &
$9.0\times10^{27}$ &  $4.0\times10^{24}$ &  $-1.29$ \\
J1140+0307 & (1)  & $-2.5$ & $1.3\times10^{-27}$  &  $4.1\times10^{-31}$ &
$2.0\times10^{28}$ &  $6.6\times10^{24}$ &  $-1.34$ \\
J1357+6525 & (1)  & $-1.8$ & $1.0\times10^{-27}$  &  $3.7\times10^{-31}$ &
$2.8\times10^{28}$ &  $1.0\times10^{25}$ &  $-1.32$ \\ 
J1434+0338 & (1)  & $-0.9$\tablenotemark{a} & $9.2\times10^{-28}$  &  $1.6\times10^{-31}$ &
$1.6\times10^{27}$ &  $2.8\times10^{23}$ &  $-1.45$ \\
  \tableline
\end{tabular}
\tablenotetext{a}{Assumed UV spectral index to calculate the flux at
  $2500{\rm \AA}$.}
\end{table*}

\section{Discussion}
We have examined X-ray/UV properties of eight AGNs with IMBHs using 12
\xmm{} observations. These constitute the first X-ray study of seven
of these AGNs excluding NGC~4395. The primary purpose of this study is
to investigate dependance of X-ray spectral and tempral properties on
BH mass as these AGNs have the lowest BH masses among all AGNs studied
in X-rays to date (see Table~\ref{t1}). All AGNs are strong X-ray/UV
sources with $0.3-10\kev$ X-ray luminosities in the range of $4.5\times10^{40} - 2.3\times10^{43}
{\rm~ergs~s^{-1}}$ and UV luminosity at $2500{\rm \AA}$ in the range
of $1.1\times10^{40} - 3.4\times10^{43}  {\rm~ergs~s^{-1}}$.

\begin{figure}
\centering
\includegraphics[width=10cm]{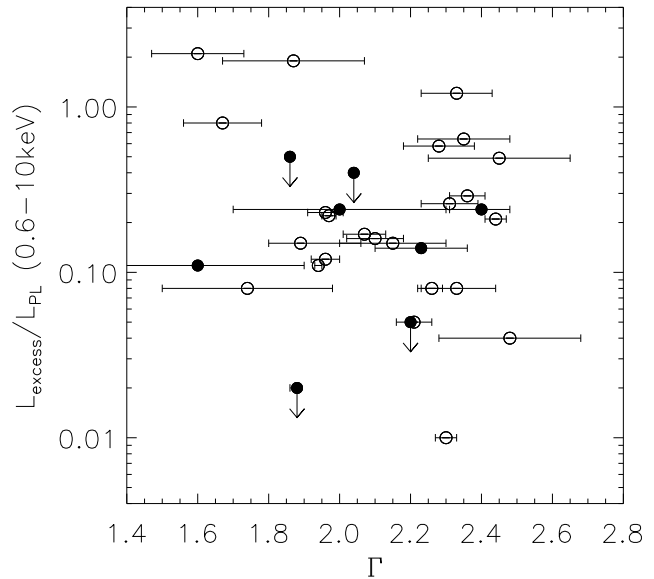}
\caption{Strength of the soft X-ray excess emission in the
  $0.6-10\kev$ band  relative to
 the power-law in the same band plotted as a function of $\Gamma$
 for AGNs with IMBH (filled circles) and NLS1s (open circles) from
  \cite{1999MNRAS.309..113V}.  The photon indices were derived from
  the best-fit models with an MCD component for all IMBH AGNs 
  except NGC~4395. }
\label{soft_excess_gamma}
\end{figure}

\subsection{X-ray continuum and soft X-ray excess emission}
The $0.3-10\kev$ primary X-ray continua of all AGNs with IMBHs
consists of either of a single power-law or a soft X-ray excess
component and a power-law. Only four AGNs show clear evidence for soft
excess emission that is well described by multicolor disk blackbody
with $kT_{in} \sim 150-200\ev$. The soft excess emission from the four
contrinute only $ltsim50\%$ in the $0.3-2\kev$ band. X-ray spectra of
three other AGNs are also consistent with the presence of soft excess
emission with similar temperatures but the component is not required
statistically. NGC~4395 is the only AGN that clearly lacks strong soft
X-ray emission.  Figure~\ref{soft_excess_gamma} compares the strength
of soft excess and the photon indices of AGNs with IMBHs and NLS1
galaxies studied by \cite{1999MNRAS.309..113V}.  The strength of the
soft excess emission is shown as the ratio of the soft excess
luminosity ($L_{excess}$) to that of the power-law component
($L_{PL}$) in the $0.6-10\kev$ band. $L_{excess}$ is the luminosity of
the MCD component in case of AGNs with IMBH, while it is the
luminosity of the blackbody component for NLS1 galaxies as
\cite{1999MNRAS.309..113V} parameterized the soft excess emission with
a blackbody component. We have plotted only the upper limits for the
three SDSS AGNs whose X-ray spectra are equally well described by a
simple PL or PL+MCD.  NLS1s show a large range in the strength of
their soft excess emission, contributing $0-68\%$ to the $0.6-10\kev$
band, while AGNs with IMBHs appear to show only weak soft excess
emission, contributing only $<35\%$ in the same energy band. NGC~4395
with the lowest $L/L_{Edd}$ has the weakest or no soft excess emission
in our sample.  Evidently, black hole mass is not the primary driver
of the strength of soft X-ray excess emission from AGNs. High
$L/L_{Edd}$ is likely the necessary condition for the strong soft
X-ray excess emission.

\begin{figure}
\centering
\includegraphics[width=8.5cm]{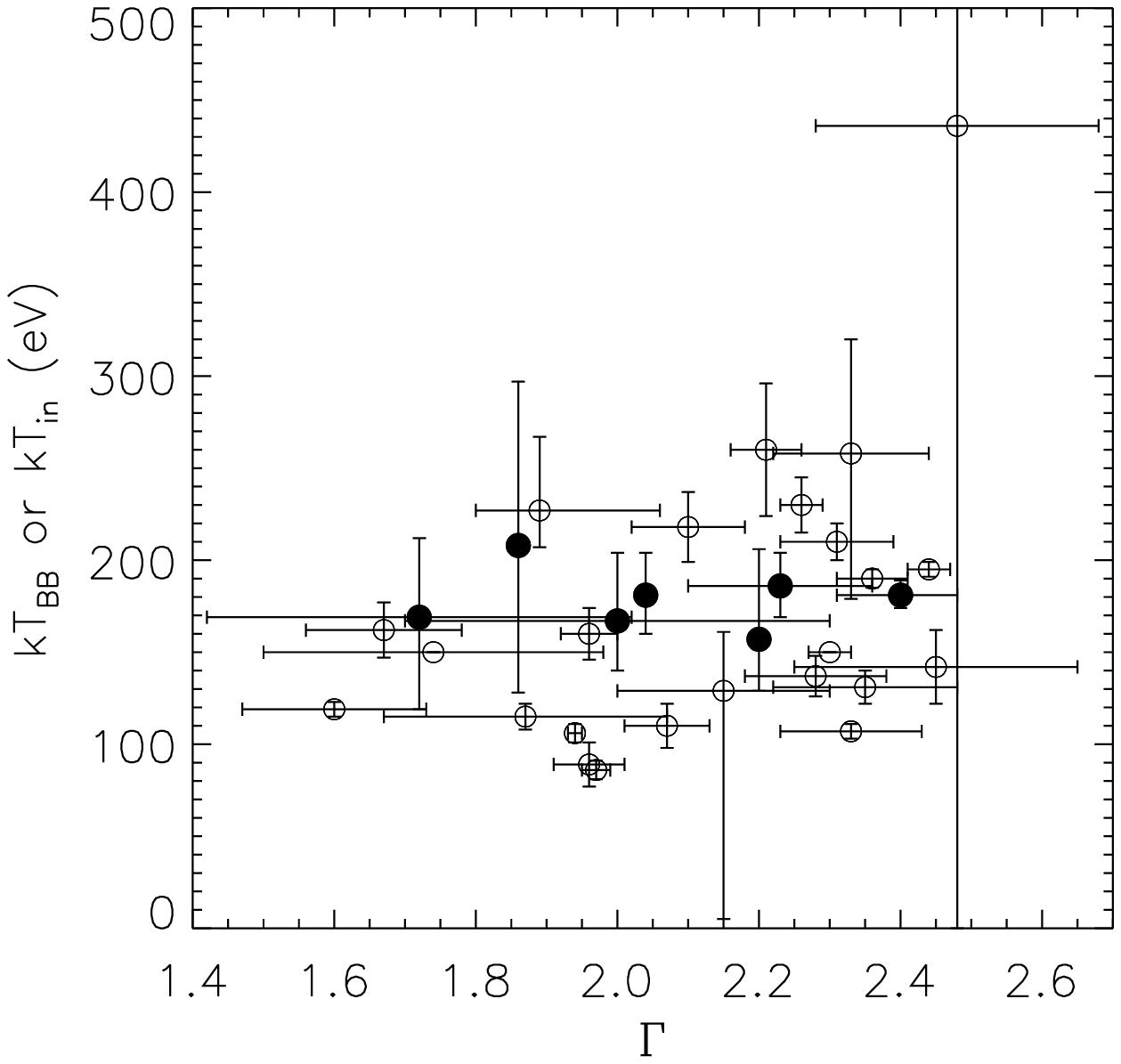}
\caption{$\Gamma$ Vs MCD $kT_{in}$ for AGNs with IMBHs (filled
  circles) or blackbody $kT_{BB}$ for NLS1 galaxies (open circles) from
Vaughan et al. (1999).}
\label{gamma_kt}
\end{figure}

The  power-law photon indices of eight AGNs range from
$\sim 1.6-2.4$ which is similar to that observed from more massive AGNs. In
particular, AGNs with IMBHs show similar range in thier photon indices
as that of NLS1 galaxies (see Fig.~\ref{soft_excess_gamma}). The
temperature of the soft excess emission detected from the AGNs
with IMBHs is also similar to that observed from NLS1 galaxies. In
Figure~\ref{gamma_kt}, we have plotted $kT_{in}$ and $\Gamma$ for
seven AGNs with IMBHs for which the data are consistent with the presence
of soft excess emission. We also show
a similar plot in Figure~\ref{gamma_kt} for 19 NLS1 galaxies studied
by \cite{1999MNRAS.309..113V} based on $0.6-10\kev$ \asca{} data. For NLS1 galaxies and AGNs with IMBHs, the blackbody temperature
appears to be uniform in the $100-300\ev$ range irrespective of the
photon index.

\subsection{X-ray Variability}
Strong X-ray variability appears to be a general property of AGNs with
IMBH, albeit at different amplitudes.  IMBH AGNs are among the most
extreme variable radio-quiet AGNs as suggested by the comparison of
the normalized excess variances in Fig.~\ref{exc_var_lum}. The excess
variances and X-ray luminosities of AGNs with the smallest black holes
are also consistent with the anticorrelation found for BLS1 and NLS1s
\citep{1997ApJ...476...70N,1999ApJS..125..297L}. In
Fig.~\ref{exc_var_lum}, IMBH AGNs and BLS1s seem to form a strong
antocorrelation and most of the dispersion in the $\sigma^2_{NXS} -
L_{2-10\kev}$ relation is caused by the NLS1 galaxies. 
%%%%%The apparent similarity in $\sigma^2_{NXS} - L_{2-10\kev}$ of
%%%%%AGNs with IMBHs with Seyfert 1s is mostly due to NGC~4395 which
%%%%%has low accretion rate. 
It
is likely that the dispersion in the $\sigma^2_{NXS} - L_{2-10\kev}$
relation is due to the variation in the accretion rates ($\dot{m_E} =
L_{bol}/L_{Edd}$) as NLS1s have high accretion rates. This trend is
indeed expected from the power density
spectra. \cite{2006Natur.444..730M} have shown a strong correlation
between the break frequency, black hole mass and the accretion rate
relative to the Eddington rate such that break time scale $T_{B}
\approx M_{BH}^{1.12}/\dot{m}_E^{0.98}$. This implies that at constant
black hole mass, the break frequency increases as $\dot{m}_E^{0.98}$,
thus increasing the integrated power over certain frequency range
covering the high frequency part of the PDS even if the PDS
normalization is the same. An increased integrated power also means
more excess variance. Thus both the lower black hole mass and higher
accretion lead to higher break frequencies and hence more excess
variance, explaining Fig.~\ref{exc_var_lum}.
From optical studies, AGNs with IMBHs show a wide range of $L/L_{Edd}$ (Table\ref{t1}). Accordingly, they also show a range in excess variance, with POX 52, having the highest $L/L_{Edd}$ also showing the highest excess variance.

\begin{figure}
\centering \includegraphics[width=9cm]{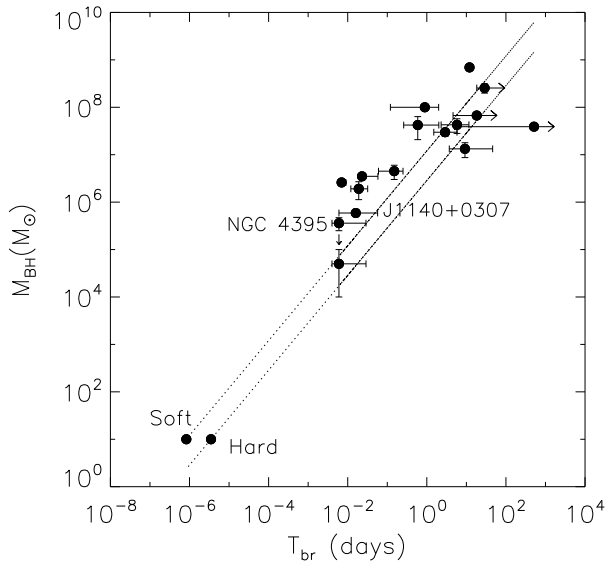}
\caption{The relation between black hole mass and PDS break time scale
  for AGNs and BHBs. The AGNs with IMBHs, NGC~4395 and
  J1140+0307 are marked. The dotted lines represent linear scaling,
  $M_{BH} \propto T_{br}$, based on the typical break times observed
  from the high/soft (shorter time scales) and low/hard states (longer
  time scales) of Cyg~X-1. See \cite{2005MNRAS.359.1469M} for details.  }
\label{mbh_tb}
\end{figure}

The power density spectra of NGC~4395, derived by
\cite{2005MNRAS.356..524V}, and J1170+0307 are consistent with a
flat power-law below a break and steep power-law above the break. The
derived break frequencies for IMBH AGNs are the highest observed among
all AGNs. The form of the PDS are broadly similar to that of more
massive AGNs. In Figure~\ref{mbh_tb}, we have plotted the break
time scales and black holes masses of the two AGNs with IMBHs along
with their massive counterparts. The break frequency and black hole
mass for NGC~4395 were taken from \cite{2005MNRAS.356..524V} and
\cite{2005ApJ...632..799P}, respectively. For massive AGNs, the data
were taken from \cite{2005MNRAS.363..586U}. The two dotted lines
represent assumed linear scaling of black hole mass with break time
scale based on the high/soft (shorter time scale at the same black
hole mass) and low/hard states of Cyg X-1
\citep[see][]{2005MNRAS.363..586U}. The data points located left to
the dotted lines represent highly accreting AGNs. The break time
scales and black hole mass of J1140+0307  is consistent
with the $M_{BH} - T_{br}$ relation for AGNs and BHBs. However,
NGC~4395 with low accretion rate ($\dot{m}_E \simeq 1.2\times10^{-3}$;
Table~\ref{t1} and Table~\ref{t7}) is inappropriately placed in
Fig.~\ref{mbh_tb}. This is due to an order of magnitude uncertainty in
the black hole mass of NGC~4395.  \cite{2003ApJ...588L..13F} argued
that the true black hole mass of NGC~4395 is in the range of $10^4 -
10^5M_{\odot}$ which is consistent with several other measurements
\citep[see][]{2005MNRAS.356..524V}. However, the black hole mass
derived from the reverberation mapping by \cite{2005ApJ...632..799P}
is an order of magnitude larger than the best guess value of
\cite{2003ApJ...588L..13F}. When this lower mass of $10^4 -
10^5M_{\odot}$ is used NGC~4395 data are consistent with other AGNs as
was mentioned by \cite{2005MNRAS.356..524V}.

With the $\sigma^2_{NXS} - L_{2-10keV}$ and $M_{BH} - T_{br}$
relations for IMBH AGNs both being consistent with that for massive
AGNs, the variability process of these low mass AGNs are intrinsically
the same to that operating in massive AGNs. The extreme variability
of AGNs with IMBHs is mainly due to their lower black hole masses.

%\subsection{Comparison with narrow-line Seyfert~1 galaxies}

%\subsection{Accretion rates and X-ray spectral states of AGNs with IMBH}
\subsection{$FWHM_{H\beta} - \Gamma_{X}$ relation}
\begin{figure*}
\centering
\includegraphics[width=8.5cm]{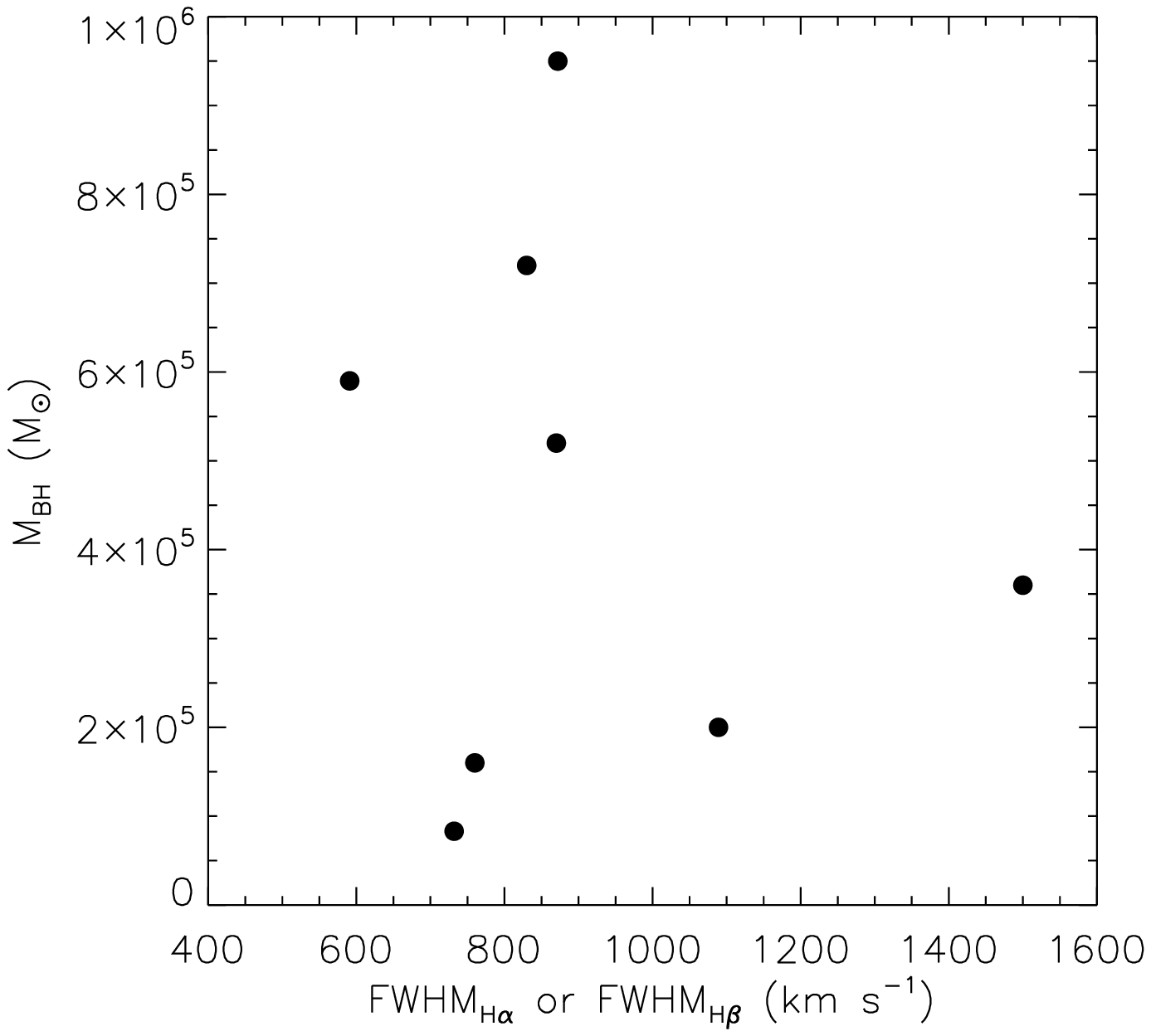}
\includegraphics[width=8.5cm]{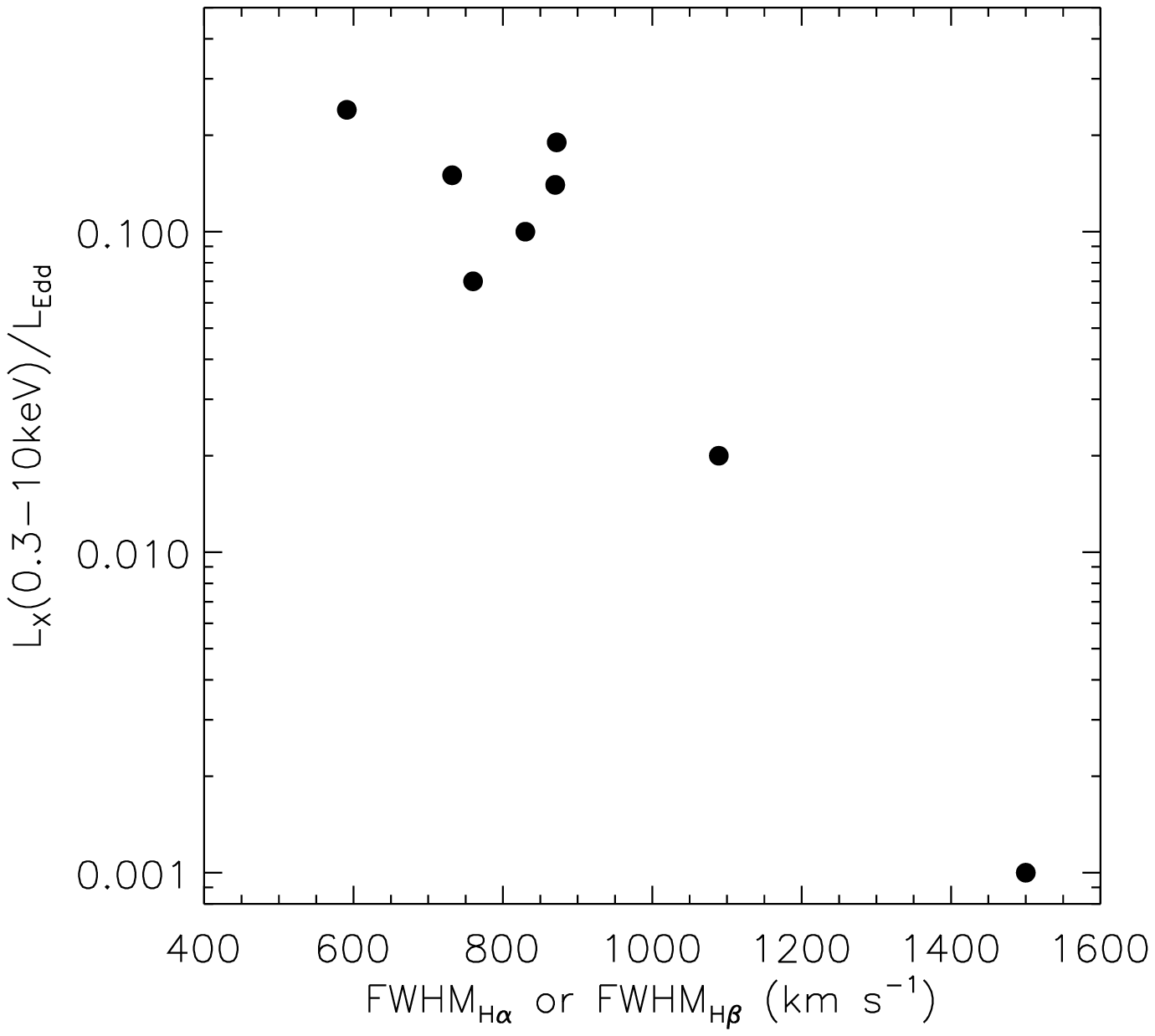}
\caption{Black hole mass ({\it left panel}) and X-ray luminosity relative to the Eddington luminosity ({\it right panel}) as a function of FWHM of the Balmer H$\alpha$ or H$\beta$ line. }
\label{hbeta}
\end{figure*}

Seyfert 1 galaxies show a large range in the width of their optical
permitted emission lines e.g., full width at half maximum (FWHM) of
the ${\rm H}\beta$ line is found to be in the range $\sim$ $1000 -
10000~{\rm ~km~s^{-1}}$. NLS1 galaxies lie at the lower end of the
line width distribution, they were originally identified with FWHM
$({\rm H}\beta) \ltsim 2000{\rm ~km~s^{-1}}$
\citep{1985ApJ...297..166O}.  By this definition, all the AGNs with
IMBH qualify as NLS1 galaxies.  NLS1 galaxies also show distinct X-ray
properties e.g., rapid variability and generally steep X-ray spectrum
\citep{1996A&A...305...53B,1999ApJS..125..297L,
  1999ApJS..125..317L,2004AJ....127..156G}. 
NLS1 galaxies show a large dispersion in their soft ($0.1-2.4\kev$)
and hard X-ray ($2-10\kev$) photon indices than the BLS1 galaxies with
larger H$\beta$ FWHM \citep{1996A&A...305...53B,1997MNRAS.285L..25B}.
The dispersion in the photon indices of IMBH AGNs is similar to that
of NLS1 galaxies. The difference is that IMBH AGNs have lower H$\beta$
FWHM on an average than that of NLS1 galaxies.
%NLS1 and BLS1 galaxies
%show an anticorrelation between the width of the optical Balmer line
%H$\beta$ and the shape of the X-ray spectrum \citep[see
%e.g.,][]{1997MNRAS.285L..25B,1999ApJS..125..317L}. However, AGNs with
%IMBHs do not follow this correlation as NGC~4395 has shown a relatively flat
%X-ray spectrum with $\Gamma \sim 1.9$ despite the FWHM of the H$\beta$
%line is narrow, $FWHM_{H\beta} \sim 1500\kms$
%\citep{1999ApJ...520..564K}. 
%
%
The current best explanation of distinct
properties of NLS1 galaxies is that they host low mass black holes
that are accreting at high fraction of their Eddington rate. In this
picture, the Balmer lines are relatively narrow due to low black hole
masses. AGNs with IMBHs play here an important role to investigate if
the narrower permitted lines arise solely due to the low black hole
mass. In Figure~\ref{hbeta}, we have plotted the black hole masses as
a function of the FWHM of the Balmer lines H$\beta$ or H$\alpha$ line
for the AGN with IMBHs.  If the black hole mass were the single
physical parameter responsible for the width of Balmers H$\beta$ or
H$\alpha$ line, we expect a strong correlation between black hole mass
and FWHM of the H$\beta$ or H$\alpha$ line. However, there is no
correlation between the two quantities.  We have used the
reverberation mass for NGC~4395 in Fig~\ref{hbeta}.  As discussed
earlier, the measured PDS break frequency for NGC~4395 is consistent
with a black hole mass in the range $10^4 - 10^5M\odot$. However, changing
the black hole mass of NGC~4395 in Fig.~\ref{hbeta} does not alter our
conclusion. In fact, NGC~4395 should have the narrowest H$\beta$ among
the AGNs with IMBH if the black hole mass were the only driving
parameter. However, H$\beta$ is broadest for NGC~4395 among the AGNs
with IMBHs (see Table~\ref{t1}). Evidently, the black hole mass alone
is not responsible for the width of permitted lines.

\begin{figure*}
\centering
\includegraphics[width=8.5cm]{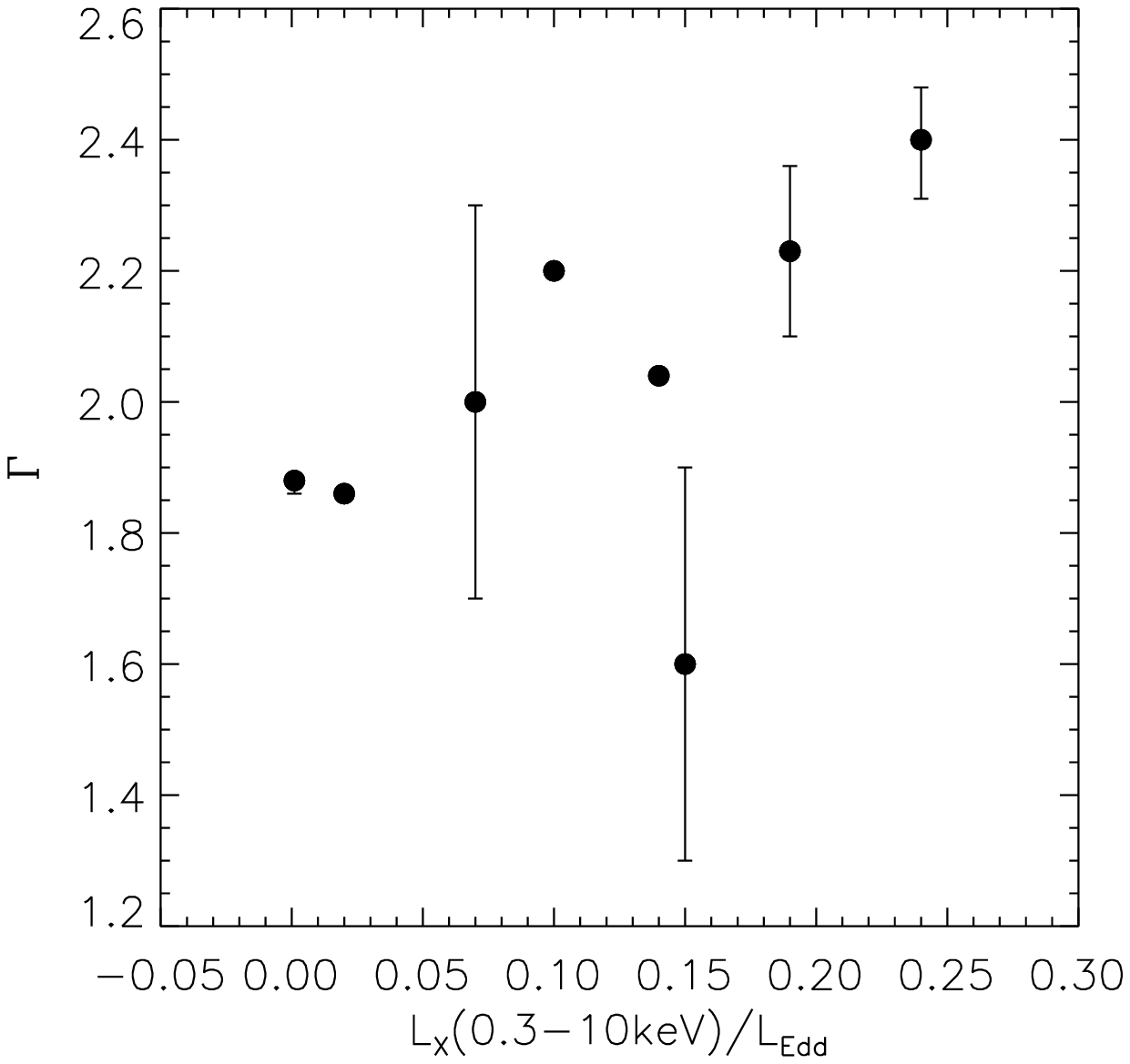}
\includegraphics[width=8.5cm]{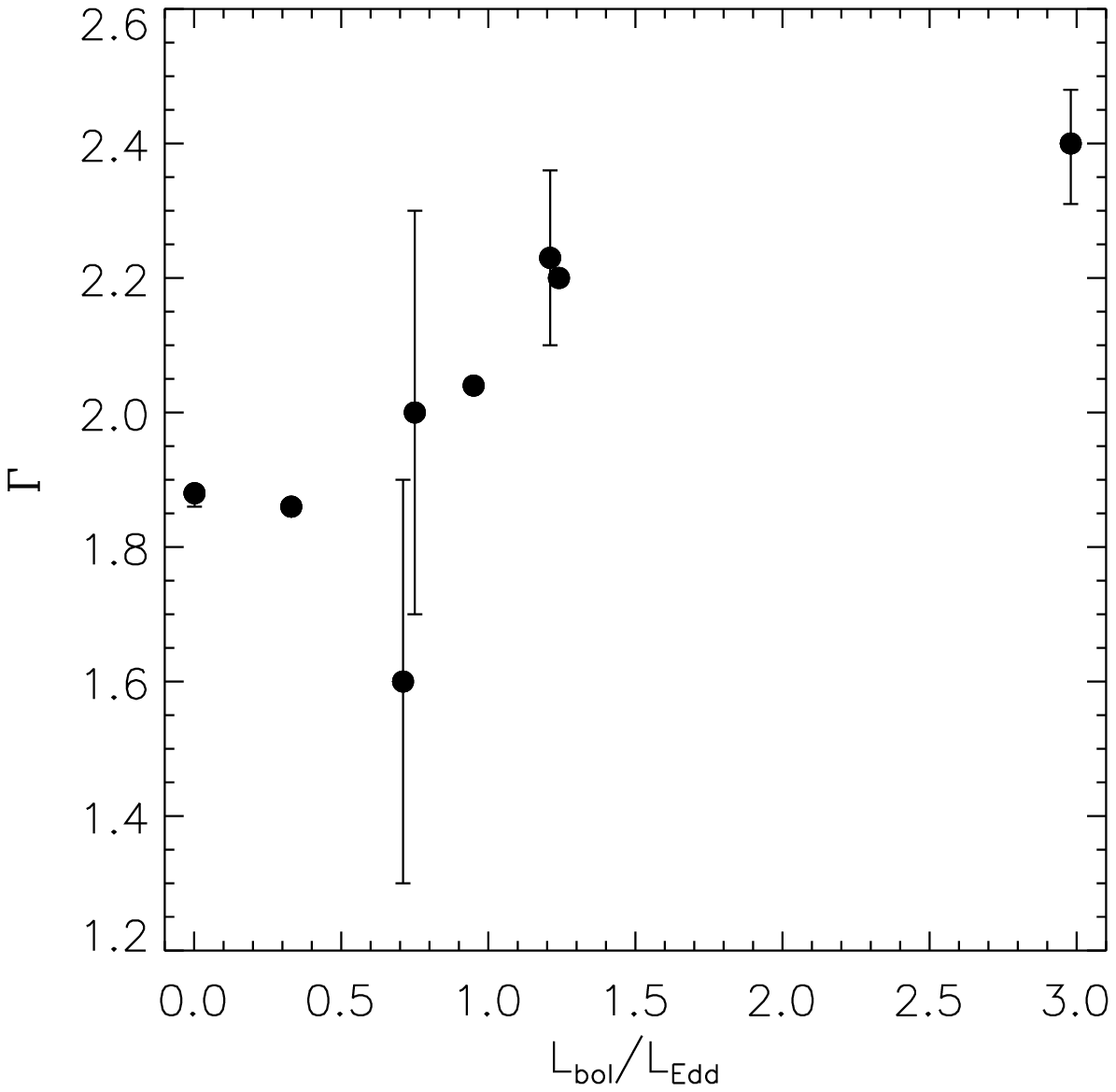}
\caption{X-ray photon index $\Gamma$ as a function of X-ray luminosity relative to the Eddington luminosity ({\it left panel}) and Eddington ratio ({\it right panel}). }
\label{gamma}
\end{figure*}

It is well known from the reverberation mapping of AGNs that the size
of the broad line region (BLR) scales with the optical luminosity
\citep[$R_{BLR} \propto L^{0.70\pm0.03}$; see
e.g.,][]{2000ApJ...533..631K}. Thus, luminosity is another parameter
in addition to the black hole mass that determines the width of the
permitted lines. This is confirmed in Figure~\ref{hbeta}, showing that
X-ray luminosity relative to the Eddington luminosity is well
correlated with the FWHM of the H$\beta$ or H$\alpha$ line. Similar
correlation is also observed between the Balmer line
widths and Eddington ratio ($L_{bol}/L_{Edd}$). The anti-correlation
between the FWHM of the H$\beta$ line and the X-ray photon index in
NLS1s and BLS1s then implies that steep X-ray spectrum is also caused
by $L/L_{Edd}$ which is a function of both black hole mass and
luminosity. However, we know that the photon index of the X-ray
spectrum does not depend on black hole mass as BHBs, massive AGNs and
AGNs with IMBHs, all show similar range of photon indices.  The photon
index, however, depends on the X-ray or bolometric luminosities
relative to the Eddington luminosity or the accretion relative to the
Eddington rate (see Figure~\ref{gamma}). Thus low black hole masses
and high absolute accretion rates resulting in high $\dot{m}_E =
L_{bol}/L_{Edd}$ are necessary criteria for AGNs to show strong
anticorrelation between the $H\beta$ line width and X-ray photon
index. Properties of AGNs with low black hole mass and low accretion
rates resulting in low $\dot{m}_E$ e.g., NGC~4395, cannot be similar
to those of NLS1 galaxies. The accretion processes in the low mass
AGNs with low $\dot{m}_E$ are likely similar to the massive AGNs with
low $\dot{m}_E$. This explains the departure of NGC~4395 from the
anticorrelation between the FWHM$_{H\beta}$ line and X-ray $\Gamma$
and similarity with BLS1s in the antocorrelation between normalized
excess variance and $2-10\kev$ luminosity. This is indeed consistent
with the fact that all NLS1s do not show the steep X-ray
spectrum. \cite{2002ApJ...581L..71D} showed NLS1s with low and high
accretion rates relative to the Eddington rate differ in their X-ray
characteristics and NLS1s with low $\dot{m}_E$ are similar to the BLS1
galaxies in terms of soft X-ray excess and shape of their X-ray
spectra. Based on \chandra{} observations of optically selected, X-ray
weak NLS1 galaxies, \cite{2004ApJ...610..737W} showed that strong soft
X-ray excess emission and steep X-ray spectrum is not a universal
characteristic of NLS1s and many NLS1s have low $\dot{m}_E$.

\subsection{Intrinsic absorption}
Among the eight AGNs with IMBHs, NGC~4395 showed the most complex
X-ray spectrum. The primary continuum of NGC~4395 observed on 2003
November is modified by multiple absorbers -- (i) fully covering
neutral absorber ($N_H \sim 5\times10^{20}{\rm~cm^{-2}}$), (ii)
partially covering neutral absorber ($N_H \sim 3.6\times
10^{23}{\rm~cm^{-2}}$ with a covering fraction of $\sim 50\%$), and
($iii$) three warm absorbers with columns in the range $N_W \sim
0.3-5.8)\times10^{21}{\rm~cm^{-2}}$ and ionization parameter $log\xi
\sim 0-3.5$).  The X-ray spectrum of POX~52 also shows evidence for a
partially covering neutral absorber ($N_H \sim 10^{23}{\rm~cm^{-2}}$,
covering fraction $\sim 90\%$) and warm absorber ($N_W \sim 7\times
10^{21}{\rm~cm^{-2}}$, $log \xi \sim 2.1$). The columns and ionization
parameters of the warm absorbers in NGC~4395 and POX~52 are well
within the range observed for Seyfert 1 galaxies studied by
\cite{2005A&A...431..111B}. Hence the presence of warm absorbers and
physical conditions do not depend on black hole mass or accretion
rate.

The dramatic spectral variability of NGC~4395 is well accounted by
variation in the absorbing components. The covering fraction of
the absorber changed from $\sim 80\%$ in 2002 May to $\sim 50\%$ in
2003 November while the absorption column increased in the same period
from $N_H = 8_{-1}^{+2}\times10^{22}{\rm~cm^{-2}}$ in 2002 May to
$3.6_{-0.3}^{+0.2}\times10^{23}{\rm~cm^{-2}}$ in 2003 November. In
addition to the higher covering fraction of the neutral absorber, the
warm absorber columns were also higher in 2002 May than 2003 November,
though the highly ionized warm absorber was not detected in 2002
May. The decreased covering fraction and warm absorber columns in 2003
November resulted almost an order of magnitude increase in the
observed $0.3-2\kev$ flux. 

Partial covering of primary X-ray source by cold matter is
not uncommon in massive AGNs. The model was first invoked in the early
days \citep{1980ApJ...241L..13H}. Recently, the partial covering or
patchy absorber has been identified in several NLS1 galaxies
\citep{2002MNRAS.329L...1B, 2004MNRAS.353.1064G, 2006AN....327.1071B,
  2004AJ....127.3161G, 2007ApJ...668L.111G}. However, the origin,
location, geometry and physical conditions of the partial covering
absorbers are not well known. Variations in partial covering
absorption has been observed on timescales of weeks to years
\citep{2004MNRAS.353.1064G, 2007ApJ...668L.111G}. In this respect,
NGC~4395 is not unusual except that it is an AGN with low accretion
rates. Hence it is unlikely that accretion disk winds act as patchy
absorber in NGC~4395 as some models invoke that a partial covering
obscuration may arise if our line of sight passes through winds that
are launched at intermediate radii of the disk
\citep[e.g.,][]{2000ApJ...545...63E,2007ApJ...661..693P}. In case of
NGC~4395, the time scale of variations is long that is apparently not
consistent with the toy model of \cite{2000A&A...356..475A} suggesting
thick clouds at $10-100$ Schwarzschild radii partially obscure the
central source. However, we note that the variability of the partial
covering absorption on short time scales, days to weeks, is
unknown. Monitoring observations of NGC~4395 with \xmm{} will be very
useful to test models for partial covering absorption.
  
\subsection{Broadband continuum properties}

\begin{figure}
\centering
\includegraphics[width=10cm]{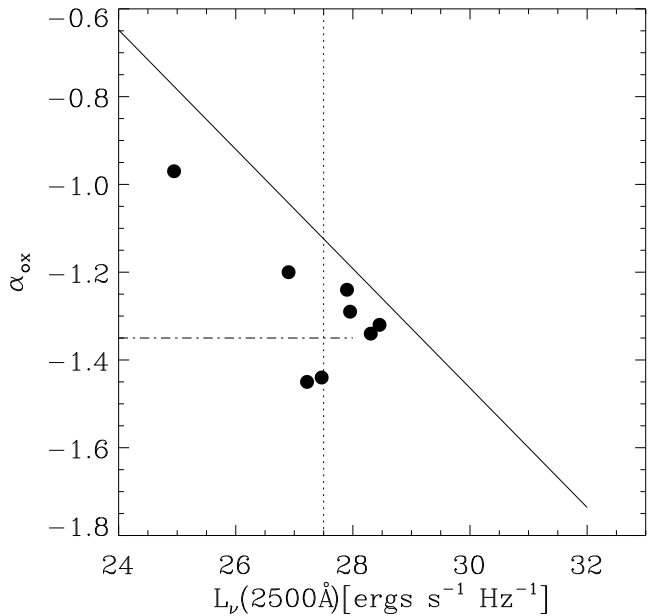}
\caption{$\alpha_{ox}$ as a function of monochromatic luminosity at
  $2500{\rm \AA}$ for AGNs with IMBHs. The solid line represents the
  primary dependance of $\alpha_{ox}$ on $log(L_{2500{\rm \AA}})$
  ($\alpha_{ox} = -0.136 logL_{UV} + 2.616$) for a large sample of 195
  AGNs \citep{2005AJ....130..387S}. The horizontal dash-dotted line
  represents the lower bounds to scatter in the $\alpha_{ox}-L_{\nu}$
  relation for the 195 AGNs. The vertocal dotted line marks the lower
  end of the $L_{\nu}(2500{\rm \AA})$ range in the large sample of
  \citet{2005AJ....130..387S} }
\label{alphaox}
\end{figure}

One of the great advantages of \xmm{} is the availability of the
optical monitor (OM), together with X-ray detectors. As a result, it
is possible to get optical-to-X-ray spectral energy distribution of
sources with simultaneous observations, thus bypassing the problems
arising due to variability. The OM fluxes of our targets are listed in
Table~\ref{tab8} while the X-ray fluxes are given in Tables~\ref{t5} and
\ref{t6}. The optical to X-ray spectral energy distribution is often
parametrized by $\alpha_{ox}$ as defined above (\S 5); the
$\alpha_{ox}$ values of our targets are listed in Table~\ref{tab9}.  It
is useful to note here that the OM observations are taken with one or
more of optical/UV filters with central wavelengths at 2070\AA,
2298\AA, 2905\AA. Thus the actual observations are taken at
wavelengths close to 2500\AA, at which the monochromatic optical flux
is calculated in the standard $\alpha_{ox}$ definition. For this
reason our measurements of $f_{\nu}$(2500\AA) are far more accurate
than those calculated from extrapolations of longer wavelengths
\citep[e.g.,][]{2005AJ....130..387S}. It is well known that AGNs are
increasingly X-ray faint relative to UV, for higher luminosities. In
Figure~\ref{alphaox} we have plotted the $\alpha_{ox}$
vs. $L_{\nu}$(2500\AA) (erg s$^{-1}$ Hz$^{-1}$). The
\citet{2005AJ....130..387S} relation for the optically selected AGNs
is shown by the solid line and our targets with IMBHs are shown as
solid circles. Ours are all low luminosity sources; all AGNs in the
Strateva et al. sample lie to the right of the vertical dotted
line. Our sample of IMBH AGNs extends the relation to lower
luminosities.  The horizontal line marks the lowest $\alpha_{ox}$ for
$\log L_{\nu} <28$ in the Strateva et al. data.  While six of our
sources appear to be consistent with the Strateva et al. relation,
SDSS J1434+0338 and POX 52 are clearly X-ray week. While broad
absorption line quasars are often undetected in X-rays, and thus
observed as X-ray faint, their intrinsic X-ray fluxes, corrected for
absorption are usually similar to non-BALQSOs;
\citealt{1996ApJ...462..637G}), X-ray weakness of the two AGNs with
IMBHs is not due to intrinsic absorption as the $\alpha_{ox}$ values
were derived after correctiong for the Galactic as well any intrinsic
neutral/warm absorption. 
% As noted earlier, the flux densities used
%to calculate $\alpha_{ox}$ were corrected for the Galactic absorption
%only. However, POX~52 has both a partial covering absorber ($N_H \sim
%8.6\times10^{22}{\rm~cm^{-2}}$, covering fraction $\sim 90\%$) and a
%warm absorber ($N_W \sim 7.1\times10^{21}{\rm~cm^{-2}}$, $log \xi \sim
%2.1$). Thus it is possible that the apparent X-ray weakness of POX~52
%could be due to strong X-ray absorption. The intrinsic $2\kev$ flux
%density corrected for the Galactic and intrinsic absorptions resulted
%in $\alpha_{ox}$ of $-1.46$ for POX~52. The optical-to-X-ray spectral
%energy distributions of POX~52 and J1434+0338 are similar. Both POX~52
%and J1434+0338 are X-ray weak compared to the rest of the six AGNs
%with IMBHs and those in the sample of \cite{2005AJ....130..387S}. Thus
%intrinsic X-ray absorption is only partially resposible for the X-ray
%weakness of POX~52. 
It is worth noting that POX~52 and J1434+0338 are
only mildly X-ray weak. NLS1 galaxies studied by Gallo et al. (2006)
show a large range $-1.76 \le \alpha_{ox} \le -0.94$. Hence the
optical-to-X-ray spectral energy distribution of all eight AGNs with
IMBHs are entirely consistent with that of NLS1 galaxies.

NGC~4395 appears to be weak in the UV and bright in X-rays as it has
largest $\alpha_{ox}$ among the IMBH AGNs. This is also evident from
the $L_{X}/L_{Edd}$ and $L_{bol}/L_{Edd}$ values. While the
contribution of $0.3-10\kev$ X-ray luminosity to the bolomtertic
luminosity is only $6-21\%$ for the six SDSS AGNs and POX~52, the
contribution is $\sim 80\%$ in the case of NGC~4395. This may suggest
different disk-corona geometries for NGC~4395 and other AGNs with
IMBHs. If the accretion disk in NGC~4395 is truncated at large inner
radii, then the disk is expected to be cooler and hence weaker
optical/UV emission. Such truncated disks are thought to exist in the
hard state of BHBs when the accretion rate is low \citep[see
e.g.,][]{2006ARA&A..44...49R}. The lack of soft X-ray excess emission,
low $L/L_{Edd}$, extremely rapid variability and the lack of UV
emission are all consistent with a truncated disk and NGC~4395 is
likely in a low state similar to the low/hard state of BHBs.

\subsection{Nature of AGNs with IMBHs}
The main objective of this study is to investigate the dependance of
X-ray properties on BH mass and thereby to identify any
departure in the accretion process that may be related to the growing
phase in the early evolution of BHs and galaxies.  As shown above,
though AGNs with IMBHs show strong X-ray variability, the excess
variance -- luminosity  and the PDS break -- BH mass relations
are consistent with the variability properties of
AGNs with SMBHs. The short break time scales are consistent with the
low BH mass of AGNs with IMBHs. The primary X-ray continuum and the
optical-to-X-ray spectral energy distributions are also similar to
that observed from massive AGNs. The propereties of neutral and warm
absorbers in NGC~4395 and POX~52 are also not distinct when compared
to that observed from Seyfert galaxies. Thus we conclude that the
observed X-ray and UV properties of AGNs with IMBHs are consistent
with the low BH mass extension of SMBHs in Seyfert galaxies and quasars. The accretion process in AGNs with IMBHs is similar to that in
more massive AGNs. There is no clear indication of any departure in
the observational characteristics that may be related to the early
evolution of black holes and galaxies if the co-evolution of BHs and
galaxies is not established in the early phases. This conclusion is
consistent with the fact that two of AGNs NGC~4395 and POX~52 follow
the same $M_{BH}- \sigma^{*}$ relation known for the massive AGNs and
normal galaxies.
\section{Summary}
We presented a systematic X-ray variability and spectral study of
eight AGNs with IMBHs based on 12 \xmm{} observations. The main
results are as follows.
\begin{enumerate}
\item Strong X-ray variability is a general property of AGNs with IMBHs. The normalized excess variances of
  AGNs with IMBHs are the largest observed among radio-quiet AGNs. The
  excess variance -- luminosity relation for AGNs with IMBHs is
  consistent with that of more massive AGNs.
\item The X-ray variability time scales of AGNs with IMBHs are the
  shortest observed from all radio-quiet AGNs, implying the most
  compact X-ray emitting regions in AGNs with IMBHs. The observed PDS
  breaks in AGNs with IMBHs are consistent with the well known $M_{BH}
  - T_{br}$ relation for AGNs and BHBs.
\item The shape of the $0.3-10\kev$ power-law continuum of AGNs with IMBHs
  is found to be in the range $\Gamma \sim 1.7 - 2.6$, entirely
  consistent with the range of phton indices observed from radio-quiet
  AGNs.
\item Only four of eight AGNs show clear evidence for soft X-ray
  excess emission similar to that observed from NLS1 galaxies. X-ray
  spectra of three other AGNs are consistent with the presence of the
  soft excess emission. NGC~4395 with the lowest $L/L_{Edd}\sim 0.001$
  is the only AGN with IMBHs that clearly lacks strong soft excess
  emission. These observations imply that
  the black
  hole mass is not the primary driver for the origin of strong soft
  X-ray excess emission. High $L/L_{Edd}$ is likely
  responsible for the strong soft X-ray excess emission from some AGNs.
\item The optical-to-X-ray spectral energy distributions of AGNs with
  IMBHs are generally consistent with that observed from narrow-line
  Seyfert 1 galaxies. 
\item Two AGNs NGC~4395 and POX~52 show evidence for multiple
  absorbers -- both neutral and warm. These absorbers are similar to
  that observed from Seyfert 1 galaxies.
\item NGC~4395 showed a remarkable X-ray spectral variability that is
  well explained by the variability in the partial covering neutral
  absorbers and the warm absorbers. 
\item Irrespective of the X-ray photon index, the width of the optical
  Balmer lines $H\alpha$ or H$\beta$ of AGNs with IMBHS are narrow and
  these AGNs do not follow the correlation between the $\Gamma$ of
  X-ray spectrum and the FWHM of the H$\beta$ line observed for NLS1
  and BLS1 galaxies. This can be understood if the primary driver of
  the width of H$\beta$ line and the shape of the primary X-ray
  continuum is not the black hole mass alone but the accretion rate
  relative to the Eddington rate.
\item The observed X-ray and UV properties of AGNs with IMBHs are
  consistent with these AGNs being the low mass extension of the
  more massive radio-quiet AGNs.
\end{enumerate}

GCD gratefully
ackowledges the support of NASA grant NNX06AE38G. This
work is based on observations obtained with \xmm{}, an ESA science
mission with instruments and contributions directly funded by ESA
Member States and the USA (NASA). This research has made use of data
obtained through the High Energy Astrophysics Science Archive Research
Center Online Service, provided by the NASA/Goddard Space Flight
Center.

\bibliography{imbh_agn}

\end{document}